%% file: template.tex
\documentclass{article}

\usepackage{arxiv}

\usepackage[utf8]{inputenc} % allow utf-8 input
\usepackage[T1]{fontenc}    % use 8-bit T1 fonts
\usepackage{hyperref}       % hyperlinks
\usepackage{url}            % simple URL typesetting
\usepackage{booktabs}       % professional-quality tables
\usepackage{amsfonts}       % blackboard math symbols
\usepackage{nicefrac}       % compact symbols for 1/2, etc.
\usepackage{microtype}      % microtypography
\usepackage{lipsum}

\title{\pkg{auditor}: an R Package for Model-Agnostic Visual Validation and Diagnostics}

\author{
  Alicja~Gosiewska\\
  Faculty of Mathematics and Information Science\\
  Warsaw University of Technology\\
  Poland\\
  \texttt{alicjagosiewska@gmail.com} \\
   \And
 Przemysław~Biecek \\
  Faculty of Mathematics and Information Science\\
  Warsaw University of Technology\\
  Faculty of Mathematics, Informatics and Mechanics\\
  University of Warsaw\\
  Poland\\
  \texttt{przemyslaw.biecek@gmail.com} \\
}

\begin{document}

\maketitle

\begin{abstract}
Machine learning models have spread to almost every area of life. They are successfully applied in biology, medicine, finance, physics, and other fields.
With modern software it is easy to train even a~complex model that fits the training data and results in high accuracy on test set. The problem arises when models fail confronted with the real-world data.

This paper describes methodology and tools for model-agnostic audit. 
Introduced techniques facilitate assessing and comparing the goodness of fit and performance of models. 
In~addition, they may be used for analysis of the similarity of residuals and for identification of~outliers and influential observations.
The examination is carried out by diagnostic scores and visual verification. 

Presented methods were implemented in the \CRANpkg{auditor} package for R. Due to flexible and~consistent grammar, it is simple to validate models of any classes.
\end{abstract}

% keywords can be removed

\input{chapters/introduction}

\newpage 

\input{chapters/related_work}

\input{chapters/architecture}

\input{chapters/model_audit}

\input{chapters/use_case}

\input{chapters/conclusion}

\section{Acknowledgements}
We would like to acknowledge Aleksandra \mbox{Grudziąż} and Mateusz Staniak for valuable discussions.
Also, we wish to thank Dr. Rafael De Andrade Moral for his assistance and help related to the \pkg{hnp} package. 

The work was supported by NCN Opus grant 2016/21/B/ST6/02176.

\bibliography{auditor}

\end{document}

%% file: chapters/introduction.tex
\section{Introduction} \label{introduction} % Introduction
\markboth{}{Introduction}
\addcontentsline{toc}{chapter}{Introduction}

%O czym jest praca? Co się w niej znajduje? Jaki jest wkład autora?

Predictive modeling is a process that uses mathematical and computational methods to forecast outcomes. Lots of algorithms in this area have been developed and are still being develop. Therefore, there are countless possible models to choose from and a lot of ways to train a new new complex model. A~poorly- or over-fitted model usually will be of no use when confronted with future data. Its~predictions will be misleading \citep{sheather2009modern} or harmful \citep{O'Neil:2016:WMD:3002861}. That is why methods that support model diagnostics are important.

Diagnostics is often carried out only by checking model assumptions. However, it is usually neglected for complex machine learning models since many of them are used as if they were assumption-free. Still, there is a need to verify their quality. 
We strongly believe that a genuine diagnosis or an~audit incorporates a broad approach to model exploration. The audit includes three objectives.
\begin{itemize}
    \item \strong{Objective 1:} Enrichment of information about model performance.
    \item \strong{Objective 2:} Identification of outliers, influential and abnormal observations.
    \item \strong{Objective 3:} Examination of other problems relating to a model by analyzing distributions of~residuals, in~particular, problems with bias, heteroscedasticity of variance and autocorrelation of residuals.
\end{itemize}

In this paper, we introduce the \CRANpkg{auditor} package for R, which is a tool for diagnostics and visual verification. 
As it focuses on residuals\footnote{Residual of an observation is the difference between the observed value and the value predicted by a model.} and does not require any additional model assumptions, most of the presented methods are model-agnostic. A consistent grammar across various tools reduces the amount of effort needed to create informative plots and makes the validation more convenient~and~available.

\begin{wrapfigure}{r}{0.5\textwidth}
\centering
      \includegraphics[width=0.49\textwidth]{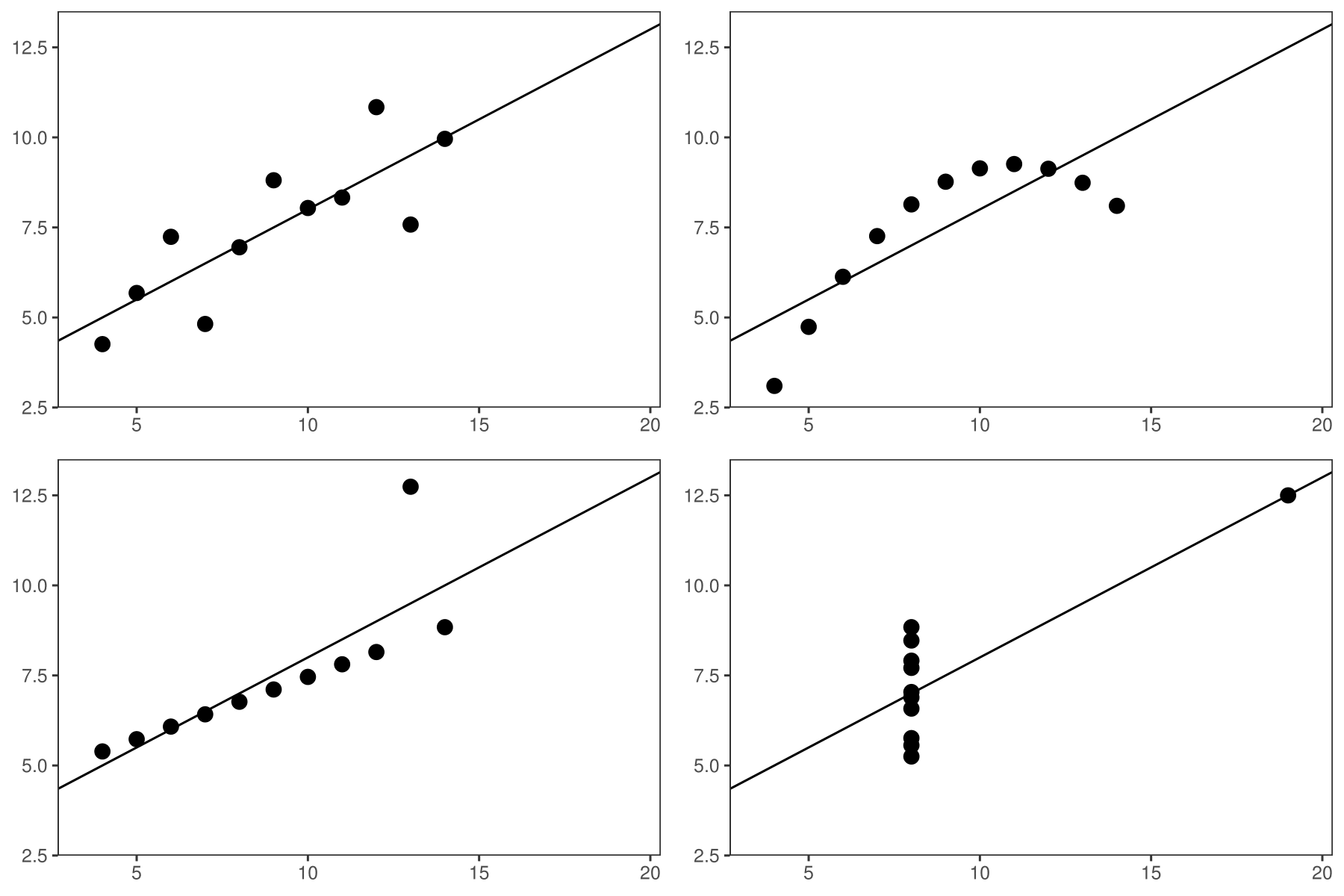}
      \caption[Anscombe Quartet]{Anscombe Quartet data sets are identical when examined with he use of simple summary statistics. The difference is noticeable after plotting the data.}
      \label{figure:anscombe}
\end{wrapfigure} 

Diagnostics methods have been a subject of much research \citep{atkinson1985plots}.  \citet{atkinson2012robust} focus on graphical methods of diagnostics regression analysis. \mbox{\citet{LIU201748}} present an overview of interactive visual model validation. 
One of the most popular tools for verification are measures of the differences between the values predicted by a~model and the observed values  \citep{doi:10.1029/JC090iC05p08995}. These tools include Root Mean Square Error (RMSE) and Mean Absolute Error (MAE) \citep{hastie01statisticallearning}. 
Such measures are used for well-researched and easily interpretable linear model as well as for complex models such as a random forest \citep{Ho:1995:RDF:844379.844681}, an XGBoost \citep{DBLP:journals/corr/ChenG16} or a neural network \citep{MASS}. 

However, no matter which measure of model performance we use, it does not reflect all aspects of the model. As indicated by \citet{breiman2001}, the linear regression model validated only on the basis of $R^2$ may lead to many false conclusions.
The best known example of this issue is the Anscombe Quartet \citep{AnscombeQuartet}. It contains four different data sets constructed to have nearly identical simple statistical properties such as mean, variance, correlation, etc. These measures directly correspond to the coefficients of the linear models. Therefore, by fitting a linear regression to the Anscombe Quartet we obtain four almost identical models (see Figure~\ref{figure:anscombe}). However, residuals of these models are very different. The Anscombe Quartet is used to highlight that the numerical measures should be supplemented by graphical data visualizations. 

The diagnostics analysis is well-researched for linear and generalized linear models. The said analysis is typically done by extracting raw, studentized, deviance, or Pearson residuals and examining residual plots. Common problems with model fit and basic diagnostics methods are presented in \citet{faraway2002practical} and \citet{Harrell:2006:RMS:1196963}
Model validation may involve both checking the overall trend in residuals and looking at residual values of individual observations \citep{littell2007sas}. 
\citet{galecki2013linear} discussed methods based on residuals for individual observation and \mbox{groups~of~observations}. 

Diagnostics methods are commonly used for linear regression \citep{faraway2004linear}. Complex models are treated as if they were assumption-free, which is why their diagnostics is often ignored. 
Considering the above, there is a need for more extensive methods and software dedicated for model auditing. Many of diagnostics tools. such as plots and statistics developed for linear models, are still useful for exploring any machine learning model. Applying the same tools to all models facilitates their comparison. 
Yet another approach to auditability of~models was proposed in \mbox{\citet{doi:10.1137/0909049}}.

The paper is organized as follows. 
Section~\ref{related work} summarizes related work and state of the art. Section~\ref{architecture}~contains an architecture of the \pkg{auditor} package. Section~\ref{audit} provides the notation.
Selected tools that help to validate models are presented in Section~\ref{audit} and conclusions can be found in Section~\ref{conclusions}.

%% file: chapters/related_work.tex
\section{Related work} \label{related work}
In this chapter, we overview common methods and tools for auditing and examining the~validity of the models. 
There are several attempts to validate. They include diagnostics for predictor variables before and after model fit, methods dedicated to specific models, and model-agnostic approaches.

%-----------------------------
\subsection{Data diagnostic before model fitting}

The problem of data diagnostics is related to the Objective 2 presented in the \nameref{introduction}, that is, the identification of problems with observations. There are several tools that address this issue. We review the most popular of them.

\begin{itemize}
\item One of the tools that supports the identification of errors in data is the \CRANpkg{dataMaid} package \citep{dataMaid}. It creates a report that contains summaries and error checks for each variable in~data. Package \CRANpkg{lumberjack} \citep{lumberjack} provides row-wise analysis. It allows for monitoring changes in data as they get processed. The~\CRANpkg{validatetools} \citep{validatetools} is~a~package for managing validation rules. 

\begin{figure}[!ht]
  \centering
  \includegraphics[width=0.7\textwidth]{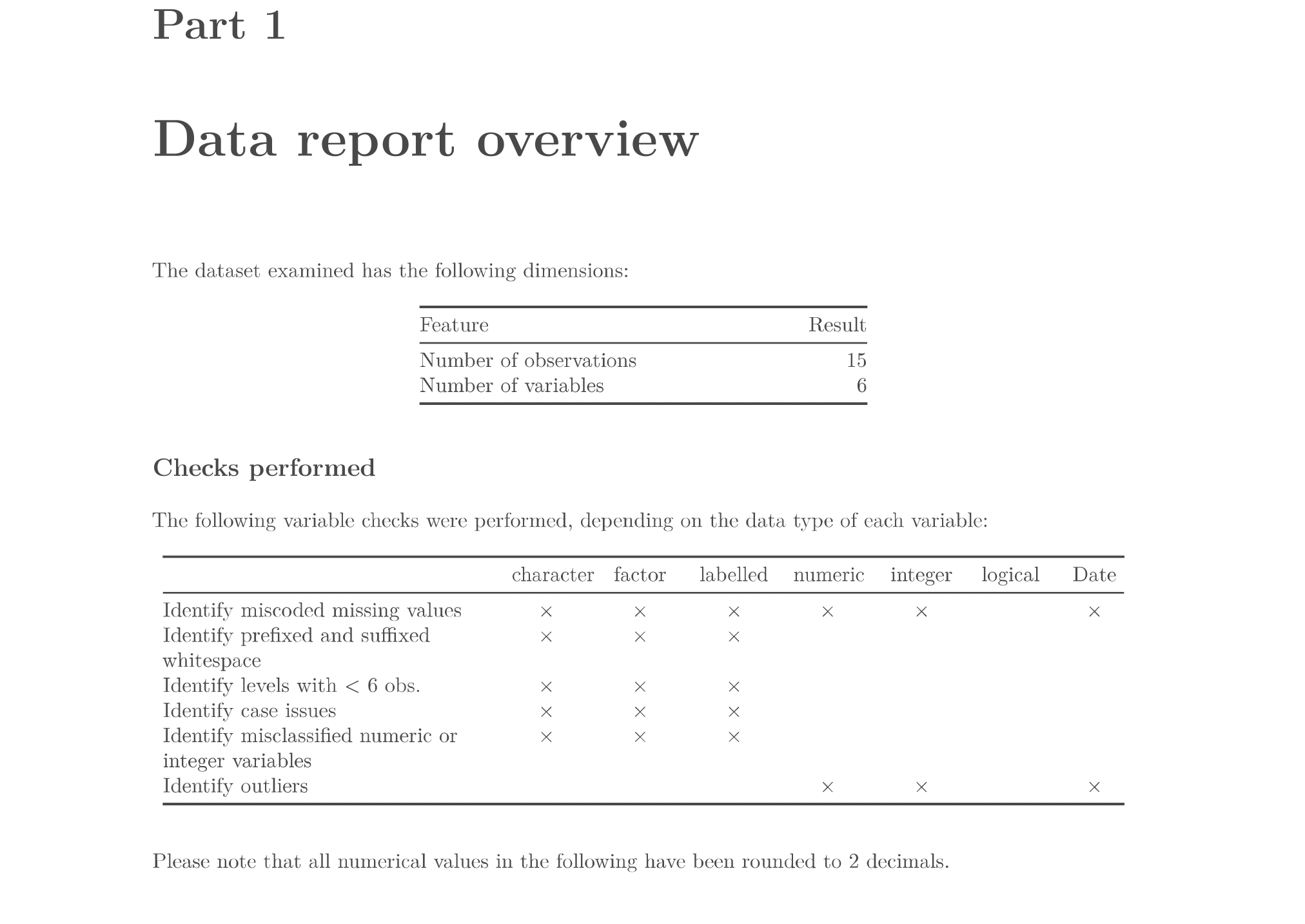}
  \caption[Exemplary page generated with dataMaid package]{Page from an exemplary report generated with \pkg{dataMaid} package. Here we have a table that summarizes checks for variables. This includes, among others, identification of missing values and outliers. In the further part of the report, there are summary tables and plots for each variable separately. The Figure comes from the report generated with the code from the \pkg{dataMaid} documentation.}
  \label{figure:DataMaid}
\end{figure}

\item  The \code{datadist} function from \CRANpkg{rms} package \citep{rms} computes distributional summaries for predictor variables. They include the overall range and certain quantiles for continuous variables, as well as distinct values for discrete variables. 
It automates the process of fitting and validating several models due to storing model summaries by~\code{datadist} function. 

\item While above packages use pipeline approaches, there are also tools that focus on specific step of data diagnostic.
The package \CRANpkg{corrgram} \citep{corrgram} calculates a correlation of variables and displays corrgrams. Corrgrams \citep{friendly:2002:EDCM} are visualizations of correlation matrices, that help to identify the relationship between variables.
\end{itemize}
\begin{figure}[!ht]
  \centering
  \includegraphics[width=0.7\textwidth]{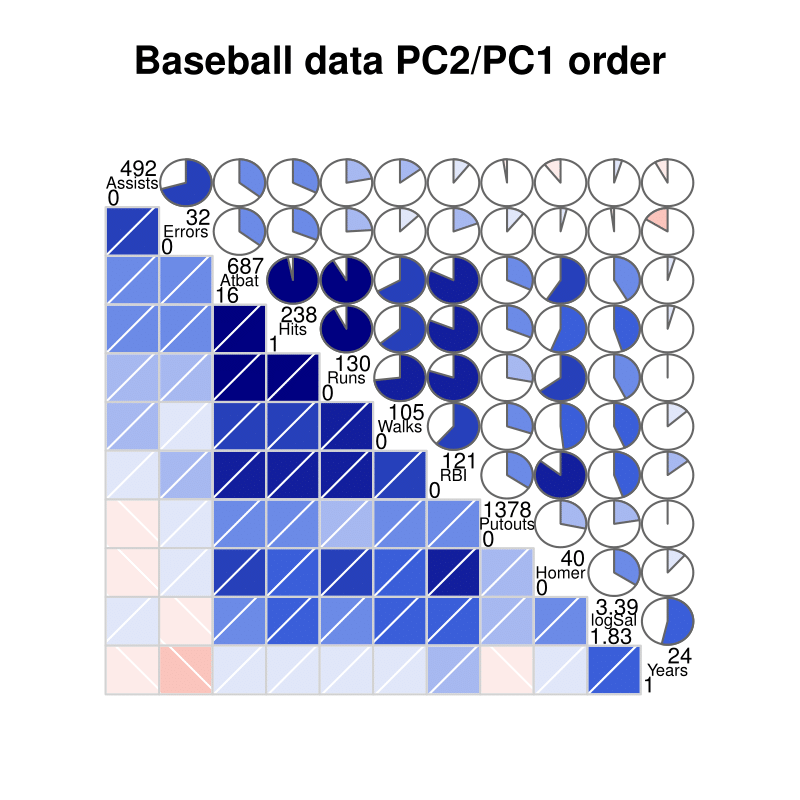}
  \caption[Corrgram plot]{A graphical display of the correlation matrix generated by the \code{corrgram} function from the \pkg{corrgram} package. Lower triangle visualizes correlations by color and intensity of shading, upper triangle by pie charts. The plot is taken from \citet{friendly:2002:EDCM} and generated with the code from vignette \href{https://cran.r-project.org/web/packages/corrgram/vignettes/corrgram_examples.html}{"Examples for the corrgram~package"}.}
  \label{figure:corrgram}
\end{figure}

%-----------------------------
\subsection{Diagnostic methods for linear models}

As linear models have a very simple structure and do not require high computational power, they have been and still are used very frequently. Therefore, there are many tools that validate different aspects of linear models. Below, we overview the most widely known tools implemented in~R~ packages.

\begin{itemize}

\item The \CRANpkg{stats} package \citep{stats} provides basic diagnostic plots for linear models. Function \code{plot} generates six types of charts for \code{"lm"} and \code{"glm"} objects, such as a plot of residuals against fitted values, a scale-location plot of $\sqrt{|residuals|}$ against fitted values and a~normal quantile-quantile plot. These visual validation tools may be addressed to the~\mbox{Objective 3} of diagnostic, related to the examination of model by analyzing the distribution of residuals.
The other three plots, that include: a~plot of Cook's distances, a~plot of residuals against leverages, and a~plot of Cook's distances against $\frac{leverage}{1-leverage}$ may be addressed to the~identification of influential observations (Objective 1). 

\begin{figure}[H]
  \centering
  \includegraphics[width=0.95\textwidth]{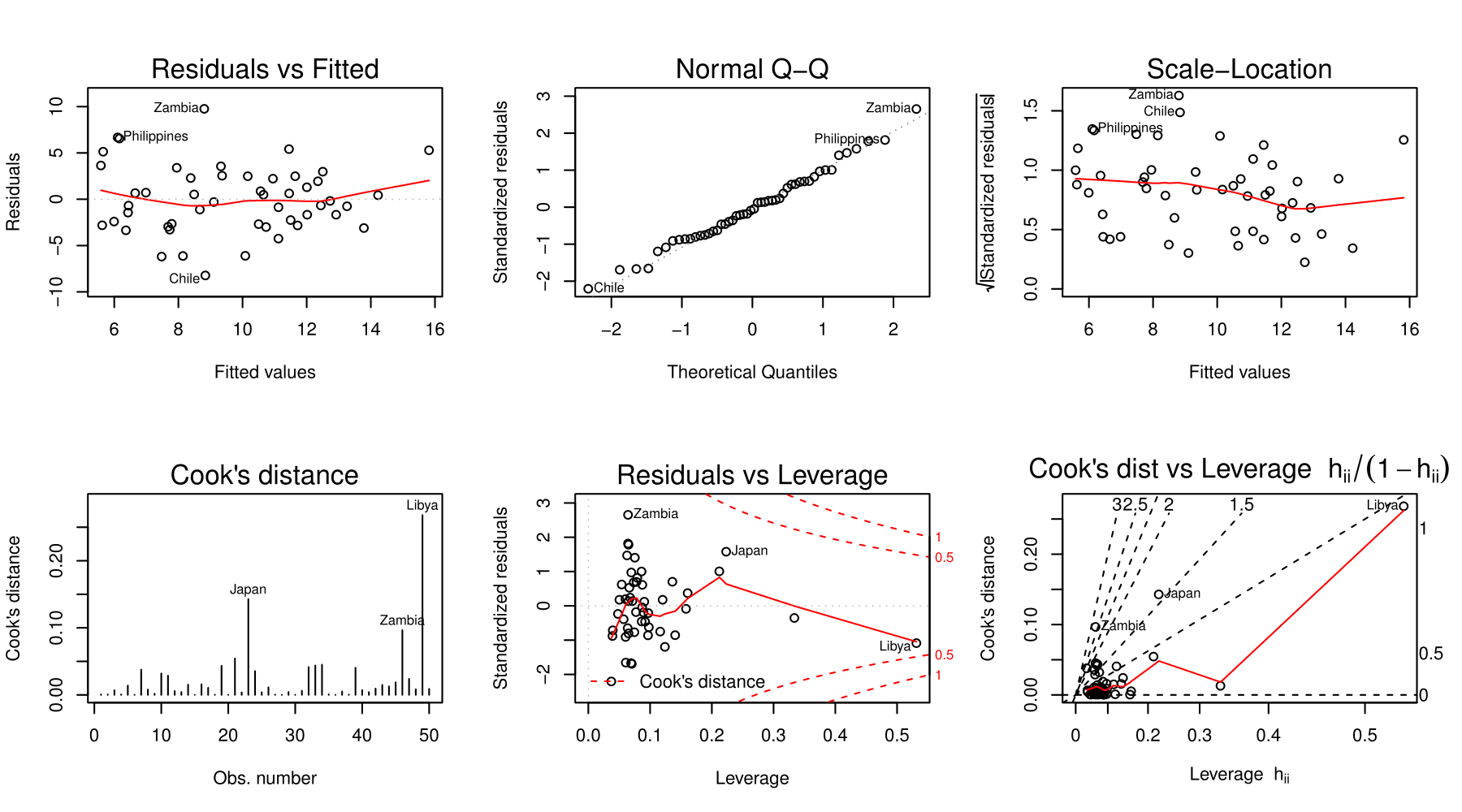}
  \caption[Diagnostic plots for linear models]{Six diagnostic plots for linear models created with generic \code{plot} function. Upper row contains plots related to the distribution of residuals, lower row is related to an influence of observations. Plots are generated with the code included in the documentation of function \code{plot.lm}.}
  \label{figure:plot.lm}
\end{figure}

\item Package \CRANpkg{car} \citep{car} extends the capabilities of \pkg{stats} by including more types of residuals, such as Pearson and deviance residuals. It is possible to plot against values of selected variables and to group residuals by levels of factor variables. What is more, \pkg{car} provides more diagnostic plots such as, among others, partial residual plot (\code{crPlot}), index plots of influence (\code{infIndexPlot}) and bubble plot of studentized residuals versus hat values (\code{influencePlot}).

These plots allow for checking both the effect of observation and the distribution of residuals, what address to the Objective 2 and \mbox{the Objective 3 respectively}.
\begin{figure}[H] 
  \centering
  \includegraphics[width=0.8\textwidth]{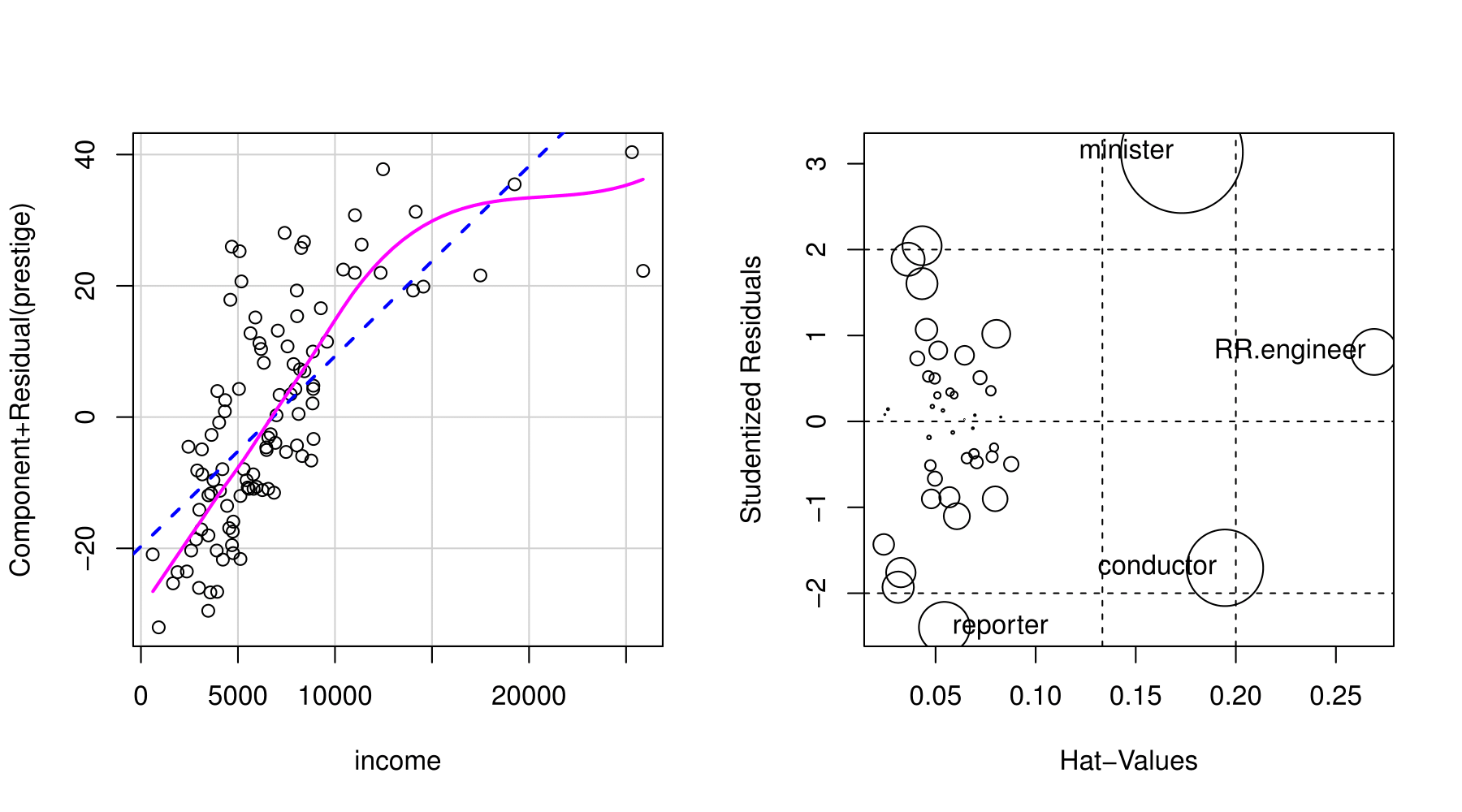}
  \caption[Diagnostic plots generated with the car package]{Two diagnostic plots generated with the \pkg{car} package. Left plot is a partial residual plot, right is a bubble plot of studentized residuals versus hat values. The areas of  circles represent Cook's distances. Plots were generated with code from the documentation of \pkg{car} package.}
  \label{figure:car}
\end{figure}

\item A linear regression model is still one of the most popular tools for data analysis due to its simple structure. Therefore, there is a rich variety of methods for checking its assumptions, for example, the normality of residual distribution and the homoscedasticity of \mbox{the variance}.

The package \CRANpkg{nortest} \citep{nortest} provides five tests for normality: the Anderson-Darling \citep{anderson1952}, the Cramer-von Mises
\citep{cramer1928composition, von1928wahrscheinlichkeit}, the Kolmogorov-Smirnov \citep{10.2307/2286009}, \mbox{the Pearson} \mbox{chi-square \citep{PearsonTest}}, and the Shapiro-Francia \citep{ShapiroFrancia} tests. 
The \CRANpkg{lmtest} package \citep{lmtest} also contains a collection of diagnostic tests: the Breusch-Pagan \citep{10.2307/1911963}, the Goldfield-Quandt \citep{doi:10.1080/01621459.1965.10480811} and the Harrison-McCabe \citep{10.2307/2286361} tests for heteroscedasticity and the~Harvey-Collier \citep{HARVEY1977103}, the Rainbow \citep{Rainbow}, and the RESET \citep{10.2307/2984219} tests for nonlinearity and misspecified functional form.
A unified approach for examining, monitoring and dating structural changes in linear regression models is provided in \CRANpkg{strucchange} package \citep{JSSv007i02}. It includes methods to fit, plot and test fluctuation processes and F-statistics.
The \CRANpkg{gvlma} implements the global procedure for testing the assumptions of the linear model (find more details in \citet{gvlma}).

The Box-Cox power transformation introduced by \citet{Box64ananalysis} is a way to transform the data to follow a normal distribution. For simple linear regression, it is often used to satisfy the assumptions of the model. Package \CRANpkg{MASS} \citep{MASS} contains functions that compute and plot profile log-likelihoods for the parameter of the Box-Cox power transformation.

\item The \CRANpkg{broom} package \citep{broom} provides summaries for about 30 classes of models. It~produces results, such as coefficients and p-values for each variable, $R^2$, adjusted $R^2$, and residual standard error.

\end{itemize}

%-----------------------------
\subsection{Other model-specific approaches}

There are also several tools to generate validation plots for time series, principal component analysis, clustering, and others.

\begin{itemize}
\item \citet{RJ-2016-060} introduced the \CRANpkg{ggfortify} interface for visualizing many popular statistical results. Plots are generated with \CRANpkg{ggplot2} \citep{ggplot2}, what makes them easy to modify.  With one function  \code{autoplot} it is possible to generate validation plots for a wide range of models. It works for, among others, \code{lm},  \code{glm}, \code{ts}, \code{glmnet}, and \code{survfit} objects.

The \CRANpkg{autoplotly} \citep{2018arXiv180308424T} package is an extension of \pkg{ggfortify} and it provides functionalities that produce plots generated by \CRANpkg{plotly} \citep{plotly}. This allows for both modification and interaction with plots.

However, \pkg{ggorftify} and \pkg{autoplotly} do not support some popular types of models, for instance, random forests from \CRANpkg{randomForest} \citep{randomForest} and \CRANpkg{ranger} \citep{ranger} packages. 

\begin{figure}[!ht]
  \centering
  \includegraphics[width=0.7\textwidth]{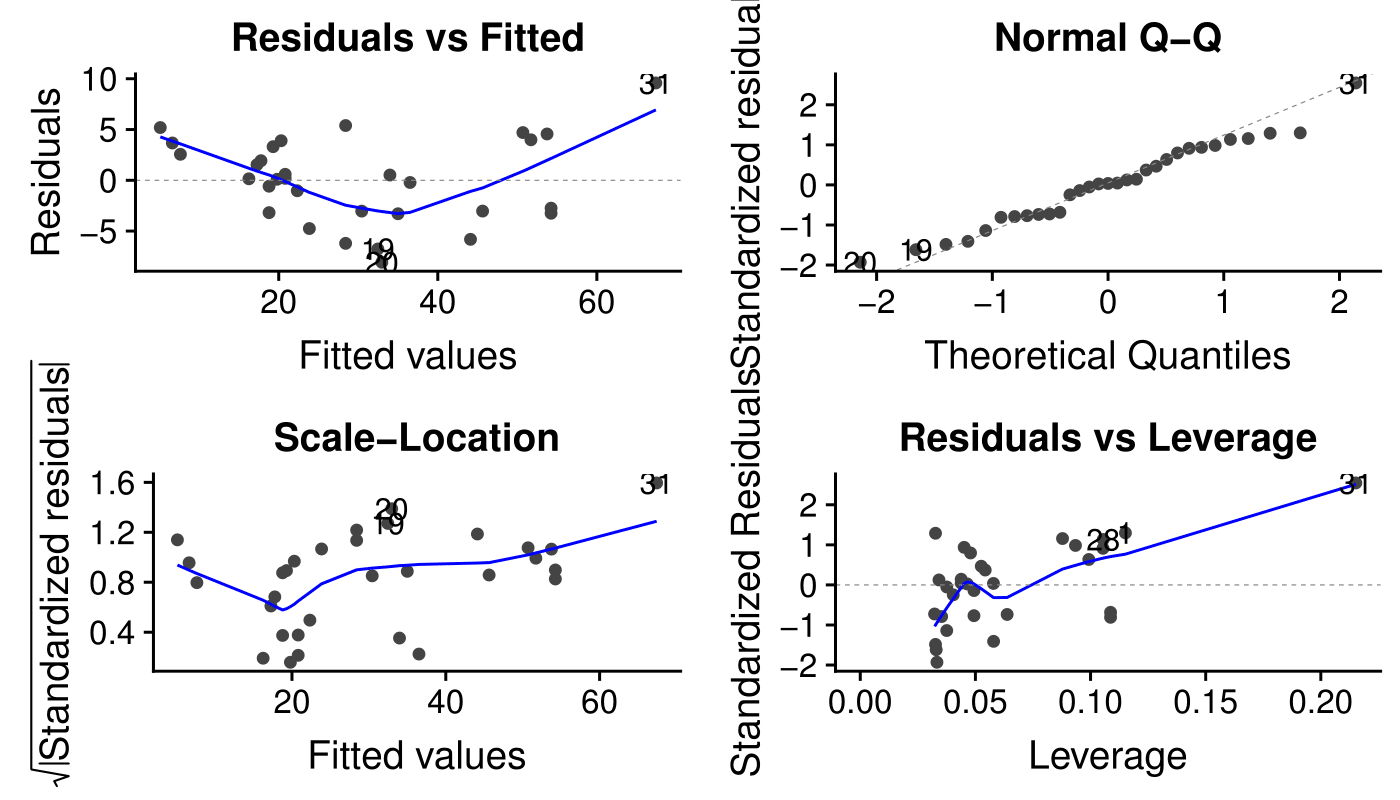}
  \caption[Diagnostic plots generated with the ggfortify package]{The diagnostic plots for linear model automaticaly generated with \code{autoplot} function from the \pkg{ggfortify} package. Plots correspond to the Objective 3 of diagnostic, that covers analysis of distribution of residuals. Plots were taken from the vignette  \href{https://cran.r-project.org/web/packages/ggfortify/vignettes/basics.html}{"Introduction to\mbox{ggfortify package}"}.}
  \label{figure:ggfortify}
\end{figure}
\item The \CRANpkg{hnp} package \citep{JSSv081i10} provides half-normal plots with simulated envelopes. 
These charts evaluate the goodness of fit of any generalized linear model and its extensions. It is a graphical method for comparing two probability distributions by plotting their quantiles against each other. 
The package offers a possibility to extend the \code{hnp} for new model classes. However, this package provides only one tool for model diagnostic.
In addition, plots are not based on \pkg{ggplot2}, what makes it difficult to modify them.

\begin{figure}[!ht]
  \centering
  \includegraphics[width=0.7\textwidth]{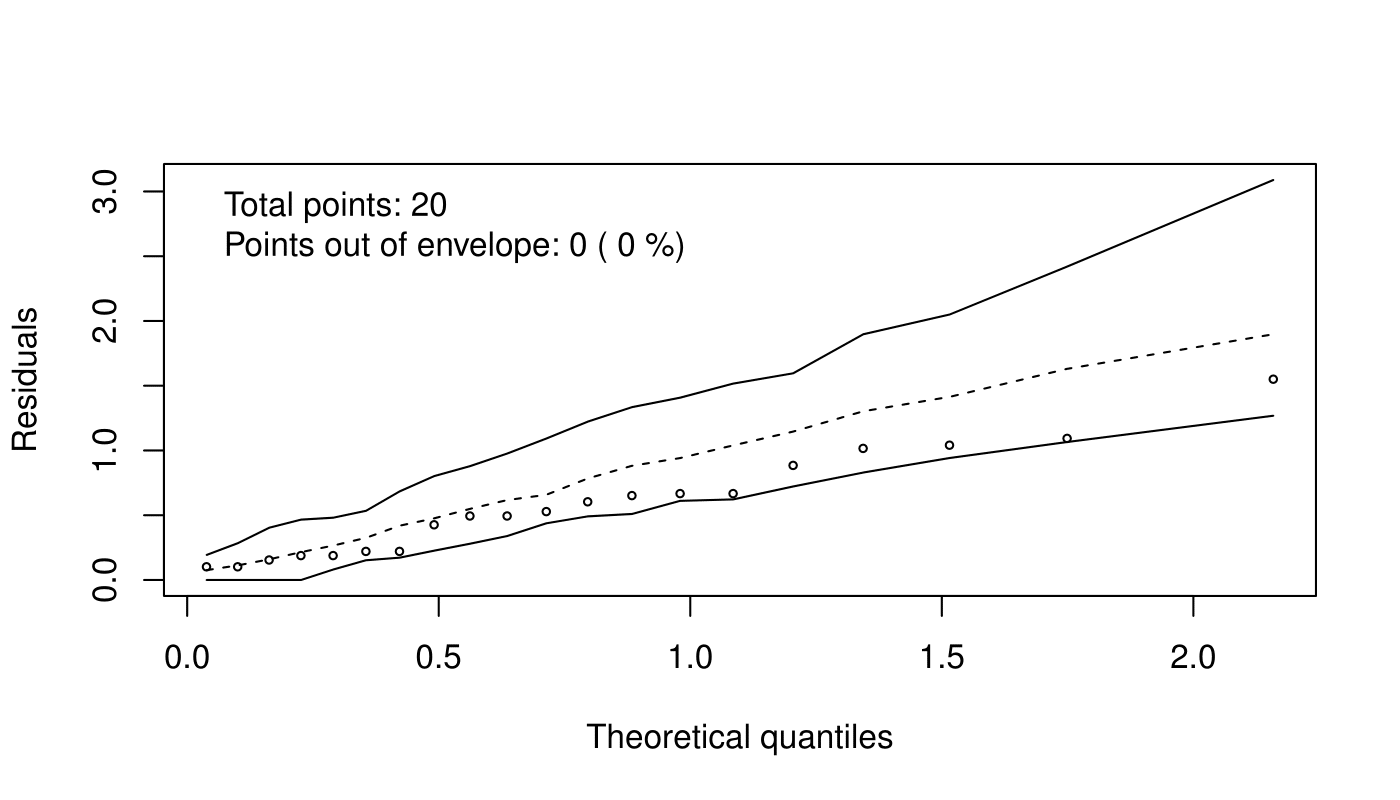}
  \caption[Half-Normal plot generated with the hnp package]{The half-normal plot generated with the \code{hnp} function from the \pkg{hnp} package. Here we have residuals and envelope that evaluate the goodness of fit.}
  \label{figure:hnp}
\end{figure}

\end{itemize}
\newpage

%-----------------------------
\subsection{Model-agnostic approach}

The tools presented above target specific model classes. The model-agnostic approach allows us to compare different~models.  
\begin{itemize}

\item The \CRANpkg{DALEX} (Descriptive mAchine Learning EXplanations) \citep{2018arXiv180608915B} is a methodology for exploration of black-box models. Main functionalities focus on understanding or proving how the input variables impact on final predictions. There are also two simple diagnostics: reversed empirical cumulative distribution function for absolute values of residuals and box plot of absolute values of residuals. 
As methods in the \pkg{DALEX} are model-agnostic, they allow for comparison of two or more models.

\begin{figure}[H]
  \centering
  \includegraphics[width=\textwidth]{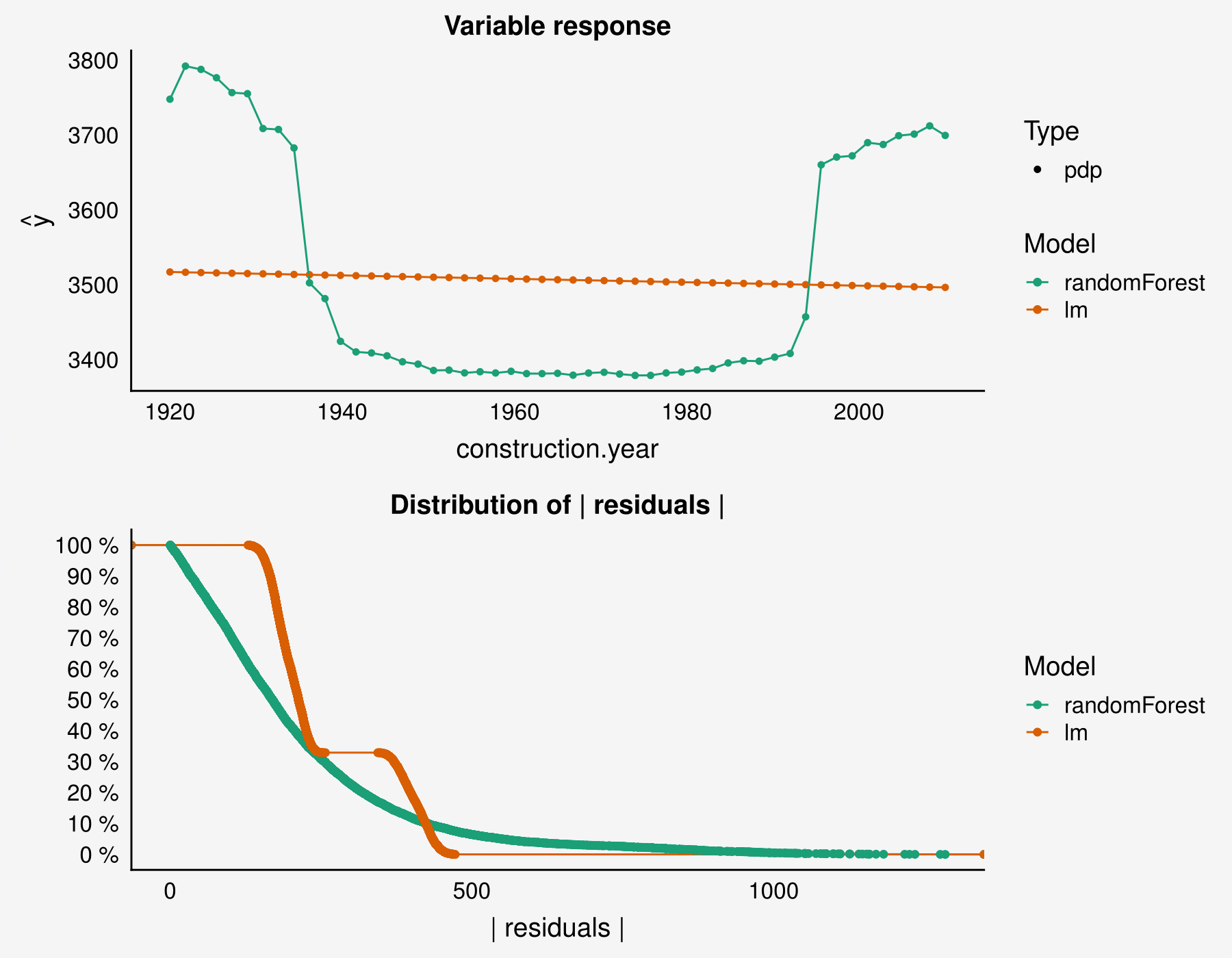}
  \caption[Plots generated with the DALEX package]{Model explanation plots generated with the \pkg{DALEX} package. The upper plot is a~Partial Dependence Plot \citep{RJ-2017-016}. The lower plot is an empirical cumulative distribution function of absolute values of residuals. Plots were taken from \citet{2018arXiv180608915B}.}
  \label{figure:DALEX}
\end{figure}

\item The package \CRANpkg{iml} \citep{iml} also contains methods for structure-agnostic exploration of model. For example, a measure of a feature’s importance by calculating the change of the model performance after permuting values of a variable.

\end{itemize}

\subsection{Model-agnostic audit}

In this paper, we present the \pkg{auditor} package for R, which fills out the part of model-agnostic validation. As~it expands methods used for linear regression, it may be used to verify any predictive model. 

%% file: chapters/architecture.tex
\section{Package Architecture}\label{architecture}

The \pkg{auditor} package works for any predictive model which returns a numeric value. It offers a~consistent grammar of model validation,  what is an efficient and convenient way to generate plots and diagnostic scores.
A diagnostic score is a number that evaluates one of the properties of a~model. That might be, for example, an accuracy of model, an independence of residuals or \mbox{an influence~of~observation.}

\begin{figure}[H]
      \includegraphics[width=\textwidth]{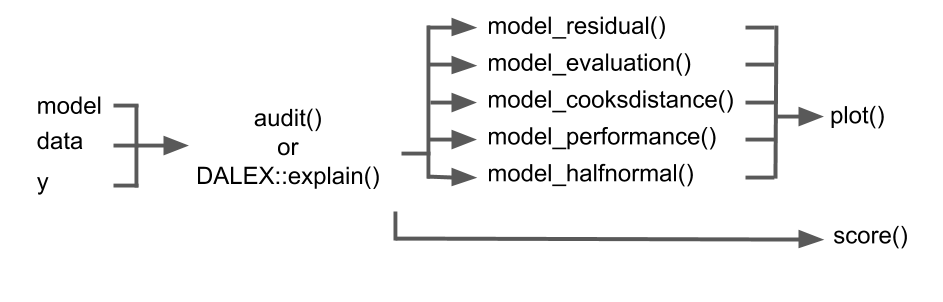}
      \caption[Architecture of the auditor]{Architecture of the \pkg{auditor}.  Function \code{explain} is implemented in the \pkg{DALEX} package. Other presented functions are implemented in the \pkg{auditor} package.}
      \label{figure:architecture}
\end{figure}

Figure~\ref{figure:architecture} presents the pipeline of the package. 
Function \code{audit} wraps up the model with meta-data, then an audited object is passed to function \code{score} or to one of the computational functions. These functions are: \code{model\_residual}, \code{model\_evaluation}, \code{model\_cooksdistance}, \code{model\_performance}, and \code{model\_halfnormal}. Results of computational functions are tidy data frames \citep{tidy-data} and then can be passed to the \code{plot} function. Once created output of the computational function contains all necessary calculations for related plots. Therefore, generating multiple plots is fast.

Implemented types of plots are presented in Table~\ref{table:functions}. Scores are presented in Table~\ref{table:scores}.
All plots are generated with \pkg{ggplot2}, what provides a convenient way to modify and combine~plots. Most of the plots have also interactive versions, they can be generated by using the function \code{plotD3} instead of the function \code{plot}.

Implemented types of plots are presented in Table~\ref{table:functions}. Scores are presented in Table~\ref{table:scores}.
All plots are generated with \pkg{ggplot2}. This provides a convenient way to modify and combine~plots.

\begin{table}[H]
    \begin{center}
        \begin{tabular}{lllcc}
            \hline
            Plot     & Function   & plot(type = ...)  &  Reg. & Class.\\ \hline
            Autocorrelation Function & \code{model\_residual} & \code{"acf"} & + & +  \\
            Autocorrelation & \code{model\_residual} & \code{"autocorrelation"} & + & + \\
            Cooks's Distances & \code{model\_cooksdistance} & \code{"cooksdistance"} & + & +  \\
            Model Correlation & \code{model\_residual} & \code{"correlation"} & + & +  \\
            Half-Normal & \code{model\_halfnormal} & \code{"halfnormal"} & + & +  \\
            LIFT Chart & \code{model\_evaluation} & \code{"lift"} &   & +  \\
            Model PCA & \code{model\_residual} & \code{"pca"} & + & +  \\
            Predicted Response  & \code{model\_residual} & \code{"performance"} & + & +  \\
            Model Ranking & \code{model\_performance} & \code{"radar"} & + & +  \\
            REC Curve & \code{model\_residual} & \code{"rec"} & + & +  \\
            Residuals & \code{model\_residual} & \code{"residual"} & + & +  \\
            Residual Boxplot & \code{model\_residual} & \code{"residual\_boxplot"} & + & +  \\
            Residual Density & \code{model\_residual} & \code{"residual\_density"} & + & +  \\
            ROC Curve & \code{model\_evaluation} & \code{"roc"} &  & +  \\
            Precision-Recall Curve & \code{model\_evaluation} & \code{"prc"} &  & + \\
            RROC Curve & \code{model\_residual} & \code{"rroc"} & + & +  \\
            Scale-Location & \code{model\_residual} & \code{"scalelocation"} & + & +  \\
           Two-sided ECDF & \code{model\_residual} & \code{"tsecdf"} & + & +  \\ \hline
        \end{tabular}
        \caption[auditor's functions generating plots]{Columns contain respectively: name of the plot, name of the computational function, value for \code{type} parameter of the function \code{plot}, indications whether the plot can be applied to regression and classification~tasks.}
        \label{table:functions}
    \end{center}
\end{table}

\begin{table}[H] 
    \begin{center}
        \begin{tabular}{lllcc}
            \hline
            Score     & Function   & score(type = ...)  &  Reg. & Class.\\ \hline
            Cook's Distance & \code{model\_cooksdistance} & \code{"cooksdistance"} & + & +  \\
            Durbin-Watson & \code{model\_residual} & \code{"dw"} & + & +  \\
            Half-Normal & \code{model\_halfnormal} & \code{"halfnormal"} & + & +  \\
            Mean Absolute Error & \code{model\_residual} & \code{"mae"} & + & +  \\
            Mean Squared Error & \code{model\_residual} & \code{"mse"} & + & +  \\
            Area Over the REC & \code{model\_residual} & \code{"rec"} & + & +  \\
            Root Mean Squared Error & \code{model\_residual} & \code{"rmse"} & + & +  \\
            Area Under the PRC & \code{model\_evaluation} & \code{"auprc"} &  & +  \\
            Area Under the ROC & \code{model\_evaluation} & \code{"roc"} &  & +  \\
            Area Over the RROC & \code{model\_residual} & \code{"rroc"} & + & +  \\
            Runs & \code{model\_residual} & \code{"runs"} & + & +  \\ 
            Peak  & \code{model\_residual} & \code{"peak"} & + & +  \\ \hline
        \end{tabular}
        \caption[auditor's functions generating scores]{Columns contain respectively: name of a score, name of a computational function, value for \code{type} parameter of function the \code{score}, indications whether the score can be applied to regression and classification~tasks.}
        \label{table:scores}
    \end{center}
\end{table}

%% file: chapters/model_audit.tex
\section{Model audit} \label{audit}

Diagnostic allows to evaluate different properties of a model. We divide them into four groups. Each of them is related to one of the following aspects and corresponding questions.
\begin{enumerate}
    \item First aspect is assessing the goodness-of-fit and whether the model does not miss relevant information (Does~the model fit data?). See subsection~\ref{model fit}.
    \item Second one is examining similarity of models (How similar models are?). See~subsection~\ref{model similarity}.
    \item Third aspect is an evaluation of model performance. (Which model has better \mbox{performance?}). See~subsection~\ref{model performance}.
    \item Last one is an identification of influential observations. (Which observations have the most impact on a model?) See subsection~\ref{oher questions}.
\end{enumerate}
These aspects are directly related to the diagnostic objectives described in the \nameref{introduction}. First two of them are related to the examination of distribution of residuals, which was proposed as the Objective 3. The third aspect is an evaluation of a model performance (Objective 1). The~last one refers to influential observations (Objective 2). 

We introduce notation to follow throughout the paper.

Let us use the following notation: $x_i = (x_i^{(1)}, x_i^{(2)}, ..., x_i^{(p)}) \in \mathcal{X} \subset \mathcal{R}^{p}$ is a vector in space $\mathcal{X}$, $y_i \in \mathcal{R}$ is an~observed response associated with $x_i$. A single observation we denote as a pair $(y_i, x_i)$ and $n$ is the~number of observations.

Let denote a model as a function $f: \mathcal{X} \to \mathcal{R}$.
Predictions of the model $f$ for particular observation we denote as
\begin{equation} 
 f(x_i) = \hat{y_i}.
\end{equation}
Row residual or, simply, residual is the difference between the observed value $y_i$ and the predicted value $\hat{y_i}$. We denote residual of particular observation as 
\begin{equation} 
 r_i = y_i - \hat{y_i}.
\end{equation}

\subsection{Aspect: Does the model fit data?} \label{model fit}

A good fitted model predicts response variable well. In this subsection, we present methods to assess the~goodness-of-fit and check whether the model does not miss any relevant information. 

To illustrate applications of the \pkg{auditor} we use an artificial data set \code{apartments} available in the \pkg{DALEX} package. First, we fit two models: simple linear regression and random forest. 
\begin{example}
  library("auditor")
  library("randomForest")

  data(apartments, package = "DALEX")
  data(apartmentsTest, package = "DALEX")

  lm_model <- lm(m2.price ~ ., data = apartments)
  set.seed(59)
  rf_model <- randomForest(m2.price ~ ., data = apartments)
\end{example}

Next step is creating \code{"modelAudit"} objects related to this two models.

\begin{example}
  lm_audit <- audit(lm_model, label = "lm", 
                data = apartmentsTest, y = apartmentsTest$m2.price)
  rf_audit <- audit(rf_model, label = "rf", 
                data = apartmentsTest, y = apartmentsTest$m2.price)
\end{example}
Below, we create objects of class \code{"modelResidual"}. They are required for generating plots.
\begin{example}
  lm_res <- model_residual(lm_audit)
  rf_res <- model_residual(rf_audit)
\end{example}

%-----------------------------------
\subsubsection{Residuals Plot}

\begin{figure}[H]
\centering
      \includegraphics[width=0.7\textwidth]{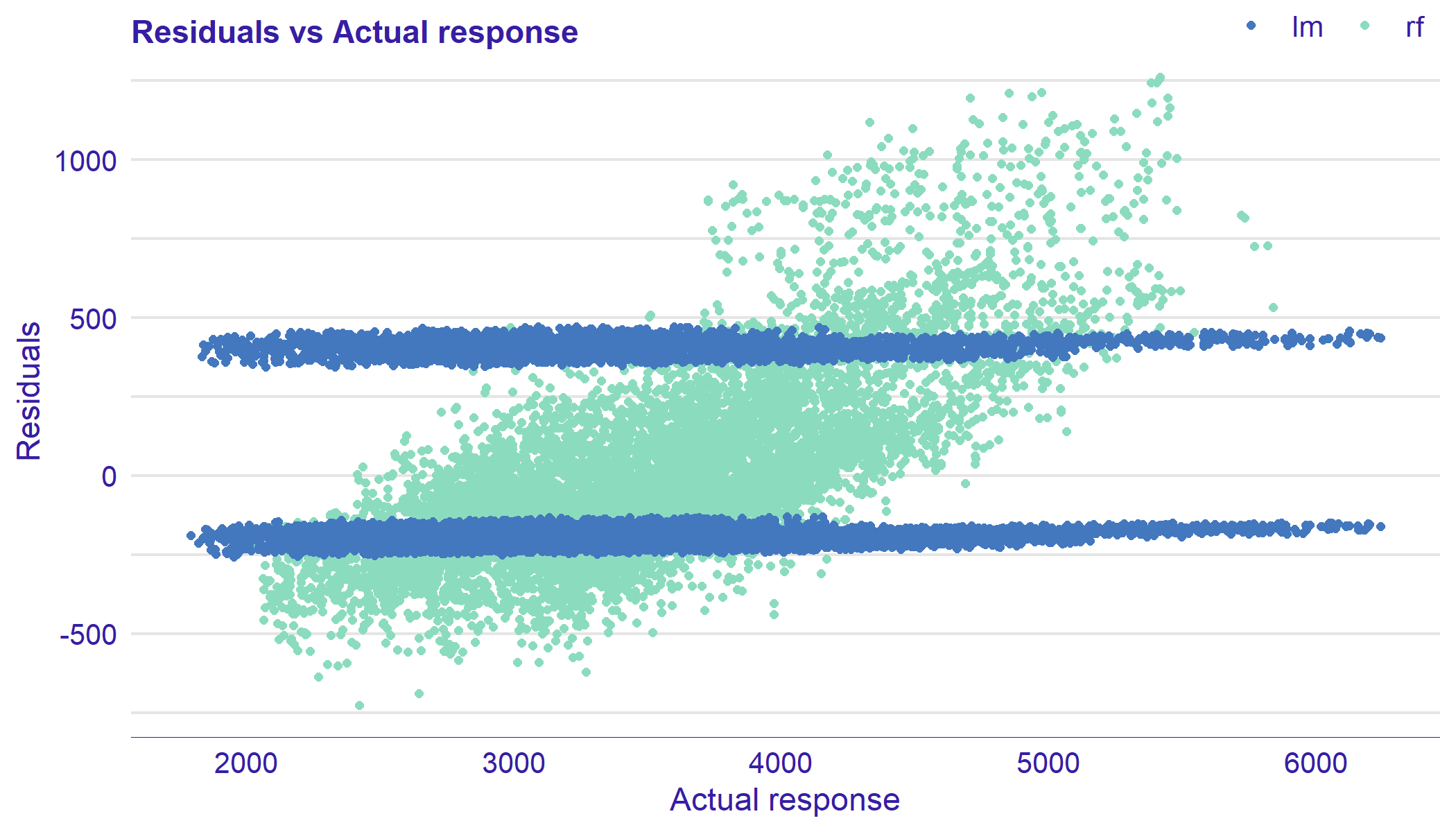}
      \caption[Residuals Plot]{Residual Plot. The patterns for both models are non-random. The values of residuals of random forest (green) increase with the increase of the fitted values. Residuals of linear model (blue) are divided into two separate groups. That suggests problems with structures of the~models.}
      \label{figure:plotResiual}
\end{figure}

Residuals vs Fitted Values is the most frequently used plot in model validation. 
It is a scatter plot of residuals $r_i$ on the y axis against fitted values $\hat{y_i}$ on the x axis. Example plot is presented in Figure~\ref{figure:plotResiual}. On alterations of this plot on the x-axis are values of the~variable.

This plot is used to detect dependence of errors, unequal error variances, and outliers.
For appropriate model, residuals should not show any functional dependency. Expected mean value should be equal $0$,  regardless of $\hat{y}$ values. Structured arrangement of points suggests a~problem with the model. It is worth looking at the observations that clearly differ from the~ others.
If points on the plot are not randomly dispersed around the horizontal axis, it may be presumed that model is not appropriate for the data.

This plot is generated by the \code{plot} function with parameter \code{type = "residual"} or by \code{plot\_residual} function. Other variants of the Residual Plot may be obtained by parameter~\code{variable} that specifies the order of residuals on the x-axis. Setting \code{variable = "\_y\_hat\_"} puts actual (true) response on the x-axis.
\begin{example}
  plot_residual(lm_res, rf_res, variable = "_y_hat_")
  # alternative:
  plot(lm_res, rf_res, type = "residual", variable = "_y_hat_")
\end{example}

%-----------------------------------
\subsubsection{Residual Boxplot Plot}

\begin{figure}[H]
\centering
      \includegraphics[width=0.7\textwidth]{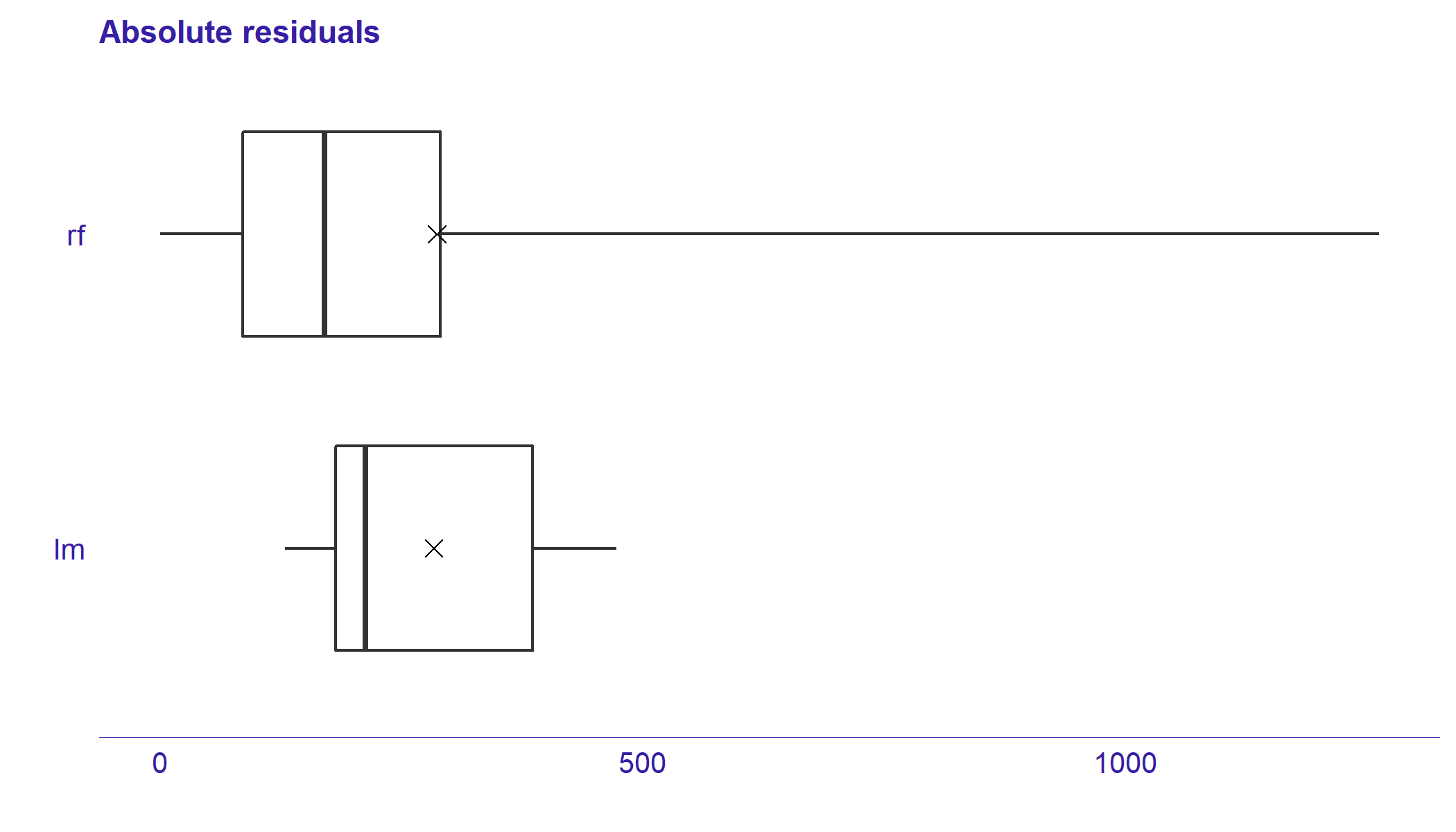}
      \caption[Residual Boxplot]{Boxplots of absolute values of residuals. Dots are in the similar places, so RMSE for both models is almost identical. However, the distribution of residuals of this two models is different. For the linear model, most of the residuals are around the average. For the random forest, most residuals are small. Nevertheless, there is also a fraction of large residuals.}
      \label{figure:plotResiualBoxplot}
\end{figure}

Residual boxplot shows the distribution of the absolute values of residuals $r_i$. They may be used for analysis and comparison of residuals. Example plots are presented in Figure~\ref{figure:plotResiualBoxplot}.

Boxplots \citep{tukey1977exploratory} usually consists of five components. Box corresponds to the first quartile, median, and third quartile. The whiskers extend to the smallest and largest values, no further than 1.5 of Interquartile Range (IQR) from the first and third quartile, respectively.  Residual boxplots consist of the sixth component. Red dot stands for Root Mean Square \mbox{Error (RMSE)}.

For the appropriate model, most of the residuals should lay near zero. A large spread of values indicates problems with a model.

This plot is generated by \code{plot\_residual\_boxplot} or by \code{plot} function with parameter \code{type = "residual\_boxplot"} function. 
\begin{example}
  plot_residual_boxplot(lm_res, rf_res)
  # alternative
  plot(lm_res, rf_res, type = "residual_boxplot")
\end{example}

%-----------------------------------
\subsubsection{Autocorrelation Plot}\label{subsection:autocorrelation}

\begin{figure}[H]
\centering
      \includegraphics[width=0.7\textwidth]{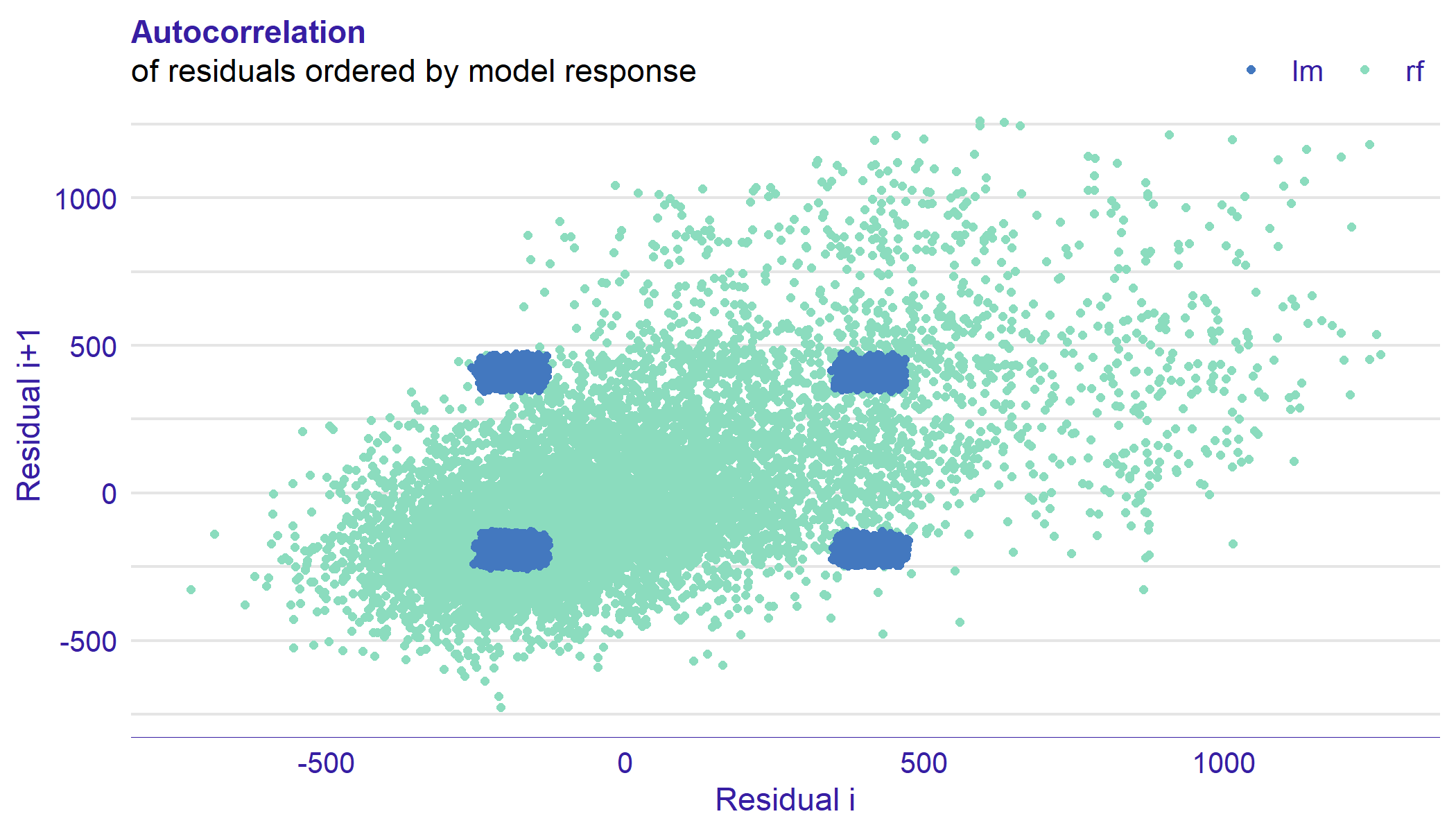}
      \caption[Autocorrelation Plot]{Autocorrelation plot of $i$-th residual vs $i+1$-th residual. The residuals of the linear model (blue) are divided into 4 groups, which indicates that they are strongly non-random. Residuals of random forest (green) show an increasing trend, which also suggests a problem with the model structure.}
      \label{figure:plotAutocorrelation}
\end{figure}

Autocorrelation Plot is a tool for checking whether there is a relationship between residuals on the lag 1. Example chart is presented in Figure~\ref{figure:plotAutocorrelation}. 
It is a scatter plot of $i$-th vs $i+1$-th residuals, ordered by fitted values or by values of one of the variables.

We expect that model residuals are independent. Therefore, points should be randomly dispersed. The~structured arrangement of residuals suggests a problem with the model.

This plot is generated by \code{plot} function with parameter \code{type = "autocorrelation"} or by function  \code{plot\_autocorrelation}. 

\begin{example}
  plot_autocorrelation(lm_res, rf_res)
  # alternative 
  plot(lm_res, rf_res, type = "autocorrelation")
\end{example}

As results of plots may be ambiguous, there are two scores implemented in the \pkg{auditor}: Durbin-Watson (DW) and Runs. They help to  asses the autocorrelation of residuals. 

\strong{The Durbin–Watson (DW)} statistic is used to detect the autocorrelation in the residuals \mbox{at lag = 1}. It is introduced by \citet{10.2307/2334313}.

\begin{definition} The Durbin-Watson test statistic is given by
  \begin{equation} \label{eq:DW}
    DW = \frac{\sum_{i=2}^n(r_i - r_{i-1})^2}{\sum_{i=1}^n r_i^2}.
  \end{equation}
  It detects only linear autocorrelation and only first order effects.
\end{definition}

We treat Durbin-Watson statistic as a score.
Its values lie between $0$ and $4$. DW around $2$ means no autocorrelation. Small value may be an evidence for positive correlation and large value for negative correlation.  \\ \\

\strong{Runs} score is the second tool for checking the independency of residuals. 

A run is a series of increasing or decreasing values. 
The $U$ is an~observed number of runs in residuals ordered by fitted values or values of one of the variables. The $\bar{U}$ is an expected number of runs and $s_U$ is the~standard deviation of the number of runs.
Runs test statistic was defined by \citet{wald1940}.
\begin{definition} The Runs test statistic is given by
\begin{equation} \label{eq:Runs}
    Z = \frac{U - \bar{U}}{s_U}.
\end{equation}
For random sequence, the distribution of $Z$ follows the $\mathcal{N}(\mu,\,\sigma^{2})$.  
\end{definition}
We consider values of Runs statistic as scores.

Both presented scores may be obtained by \code{score} function with \code{type = "dw"} or \mbox{\code{type = "runs"}.}
\begin{example}
  score_dw(lm_audit)
  score_dw(rf_audit)
  score_runs(lm_audit)
  score_runs(rf_audit)
\end{example}

%-----------------------------------
\subsubsection{Autocorrelation Function Plot}

\begin{figure}[H]
  \centering
      \includegraphics[width=0.9\textwidth]{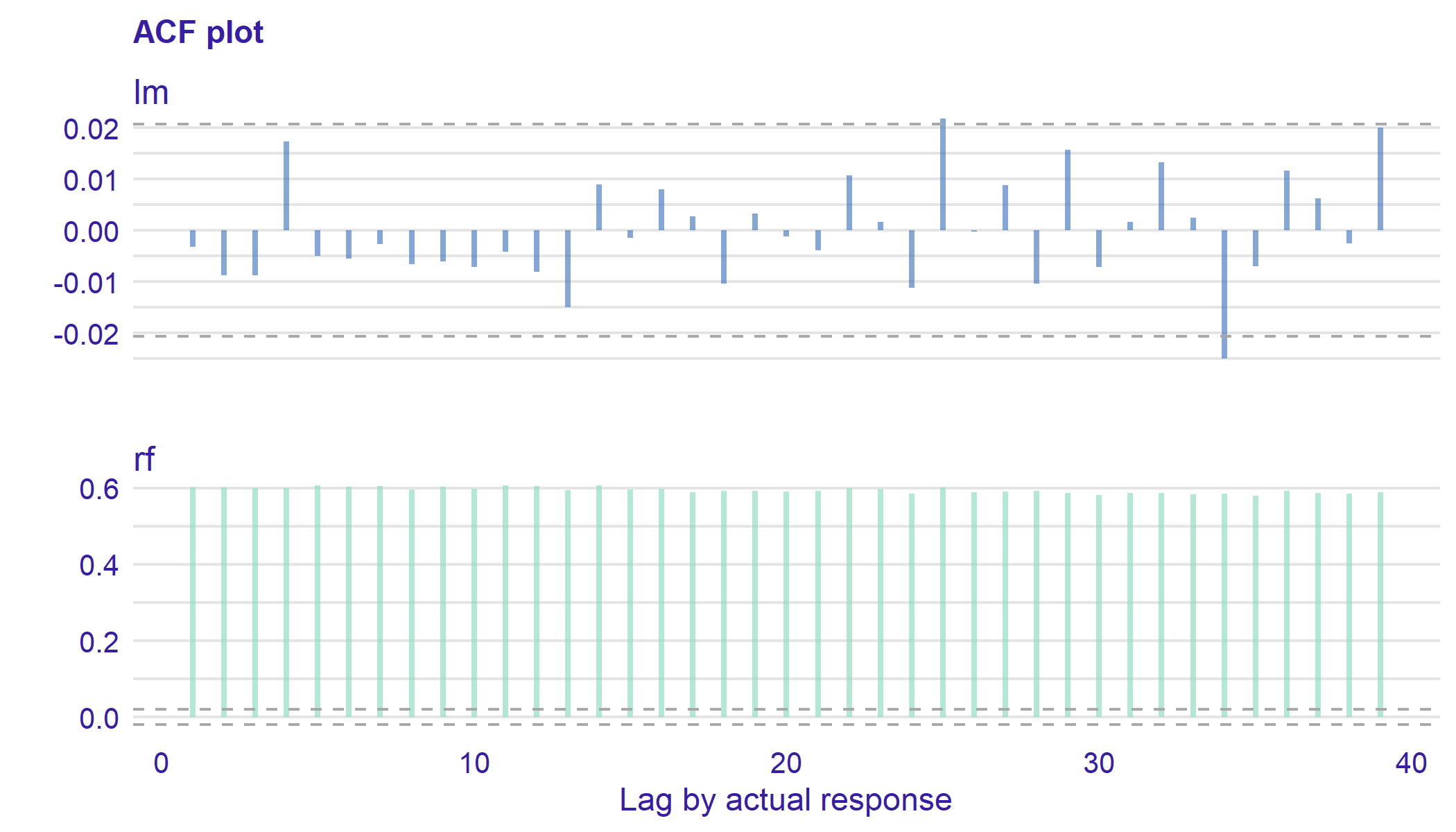}
      \caption[Autocorrelation Function Plot]{Autocorrelation Function plot. All values for random forest (blue) go out of the confidence intervals. That suggests a problem with the correlation of residuals. Residuals of the~linear model (green) do not indicate serious correlation problems.}
      \label{figure:plotACF}
\end{figure}

Autocorrelation Function (ACF) is a tool for finding patterns in the data. In particular, the~correlation between observations. Figure~\ref{figure:plotACF} shows an example of the Autocorrelation Function Plot of residuals. 

There are values of ACF on the y axis against lags on the x axis. As the value of autocorrelation of lag $0$~always equals $1$, it is skipped. Blue horizontal lines are confidence intervals.

We assume that $(X_1, X_2, ... X_n) \in \mathbb{R}^n$ are observations of a time series $X_t$ with mean $\mu$.
Covariance function $\gamma_t = Cov(X_{t+ \tau }, X_{\tau} )$ and correlation function $\rho_t = Cov(X_{t+\tau},X_{\tau})$ do not depend on $\tau$.
The following estimators come from \citet{venables2013modern}.
\begin{definition}
  The sample autocovariance function is given by
  \begin{equation}
    \hat{\gamma}(t) = \frac{1}{n}\sum_{s=max{(1,-ht)}}^{min{(n-t,n)}}(X_{s+t}-\bar{S})(X_s-\bar{X})
  \end{equation}
  and the sample autocorrelation function is given by
  \begin{equation}
    \hat{\rho}(t) = \frac{\hat{\gamma}(t)}{\hat{\gamma}(0)}.
  \end{equation} 
\end{definition}

Now, let us consider residuals $r_i$ as elements of the time series $X_t$. Autocorrelation Function Plot visualizes the sequence of $(\hat{\rho}(t))$.

This plot is generated by \code{plot} function with parameter \code{type = "acf"} or by function  \code{plot\_acf}. 

\begin{example}
  plot_acf(lm_res, rf_res, variable = "_y_hat_")
  # alternative
  plot(lm_res, rf_res, type = "acf", variable = "_y_hat_")
\end{example}

%-----------------------------------
\subsubsection{Scale-Location Plot}\label{subsection:scalelocation}

\begin{figure}[H]
      \includegraphics[width=\textwidth]{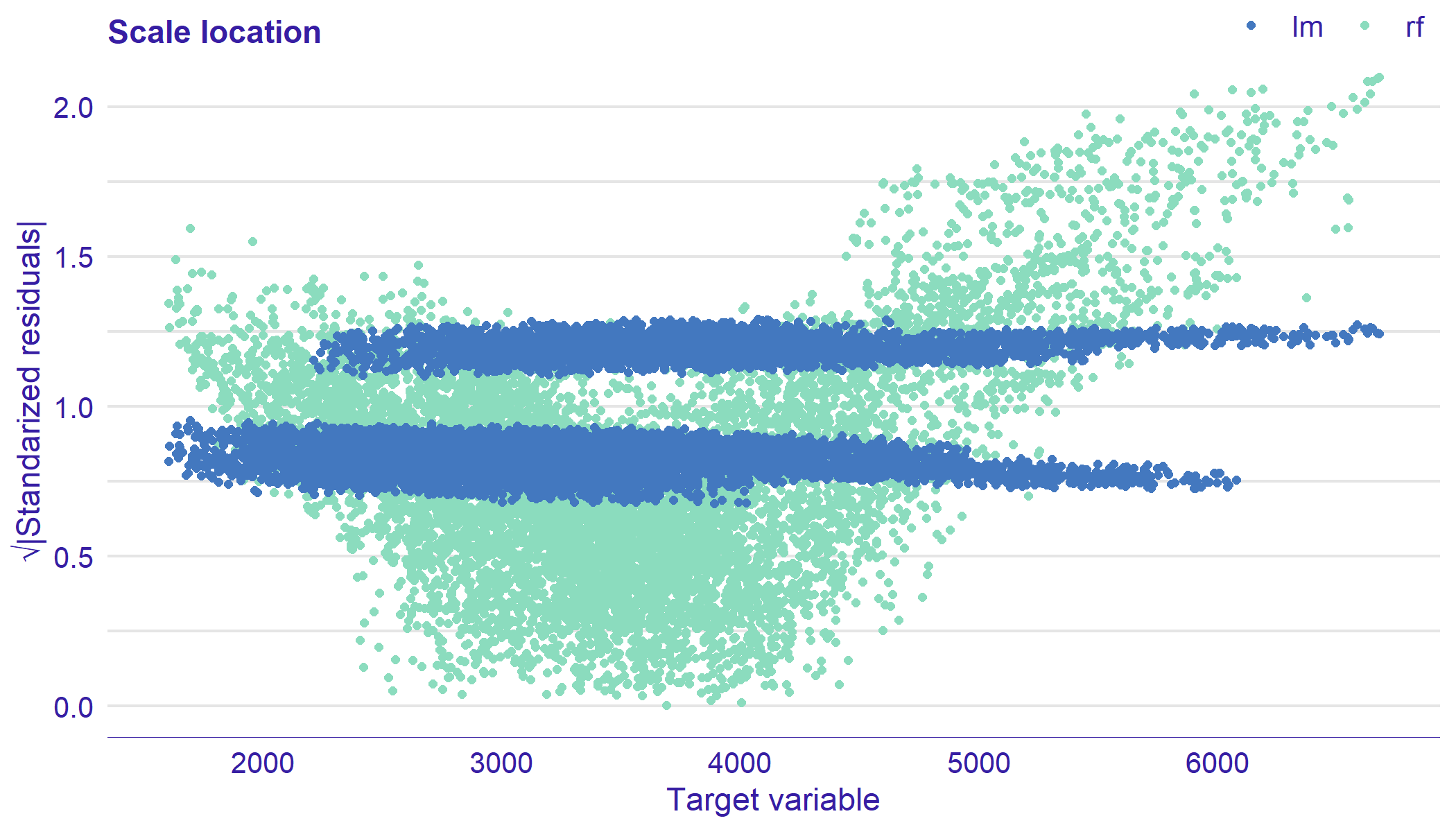}
      \caption[Scale Location Plot]{Scale Location plot. The trend in residuals of random forest (green) suggests that variance of residuals changes with the change in fitted values. The linear model (blue) seems to have no problems with variance. However, residues formed into separate groups suggest a~problem with model structure.}
      \label{figure:plotScaleLocation}
\end{figure}

Scale Location Plot is used to visualize the variance of the residuals. Example chart is presented in Figure~\ref{figure:plotScaleLocation}. This plot is similar to the Residuals Plot, but it uses the square root of the standardized residuals instead of~raw residuals. 
Let $\bar{r}$ be  mean of residuals,
\begin{equation}
    \bar{r} = \frac{1}{n} \sum_{i=1}^n r_i .
\end{equation}

\begin{definition}
  The standardized residual $r_i^{std}$ is the $i$-th residual divided by the residual standard deviation:
  \begin{equation}
    r_i^{std} = \frac{r_i}{\frac{1}{n-1}\sum_{i=1}^n(r_i - \bar{r})^2}. 
  \end{equation}
\end{definition}

On the x axis, there are responses fitted by a model. On the y axis, there are square roots of the absolute value of the standardized residuals. Blue line is an estimate of the conditional mean function. Black dots represent peaks explained below.
An alternative version of this plot contains values of one of the variables on the x axis.

The presence of any trend suggests that the variance depends on fitted values, which is against the~assumption of homoscedasticity. 
Not every model has an explicit assumption of homogeneous variance, however, 
the heteroscedasticity may indicates potential problems with the goodness-of-fit.

This plot is generated by \code{plot} function with parameter \code{type = "scalelocation"} or by function  \code{plot\_scalelocation}. Variable on the x axis may be specified by \code{variable} parameter.

\begin{example}
  plot_scalelocation(lm_res, rf_res)
  # alternative
  plot(lm_res, rf_res, type = "scalelocation")
\end{example}

A tool for assesing the homoscedasticity of variance is \strong{Peak} test introduced by \citet{doi:10.1080/01621459.1965.10480811}. 
A peak is an occurrence of observation $i$ such that $|r_i| \geq |r_j|$ for all $i > j$.
\begin{definition}
  We define Peak score as
  \begin{equation}
    P = \frac{1}{n}  \sum_{i=1}^n \mathbb{1} \{ \forall_{j < i}: |r_j| \leq |r_i| \}.
  \end{equation}
\end{definition}
Let us note that $P \in (0,1]$. If the residuals $r_i$ are heteroscedastic, the Peak score is close~to~$1$.

Scores may be obtained by \code{score} function with \code{type = "peak"} or \code{score\_peak} function.

\begin{example}
  score_peak(lm_audit)
  score_peak(rf_audit)
\end{example}

%-----------------------------------
\subsubsection{Residual Density Plot}

\begin{figure}[H]
\centering
      \includegraphics[width=0.75\textwidth]{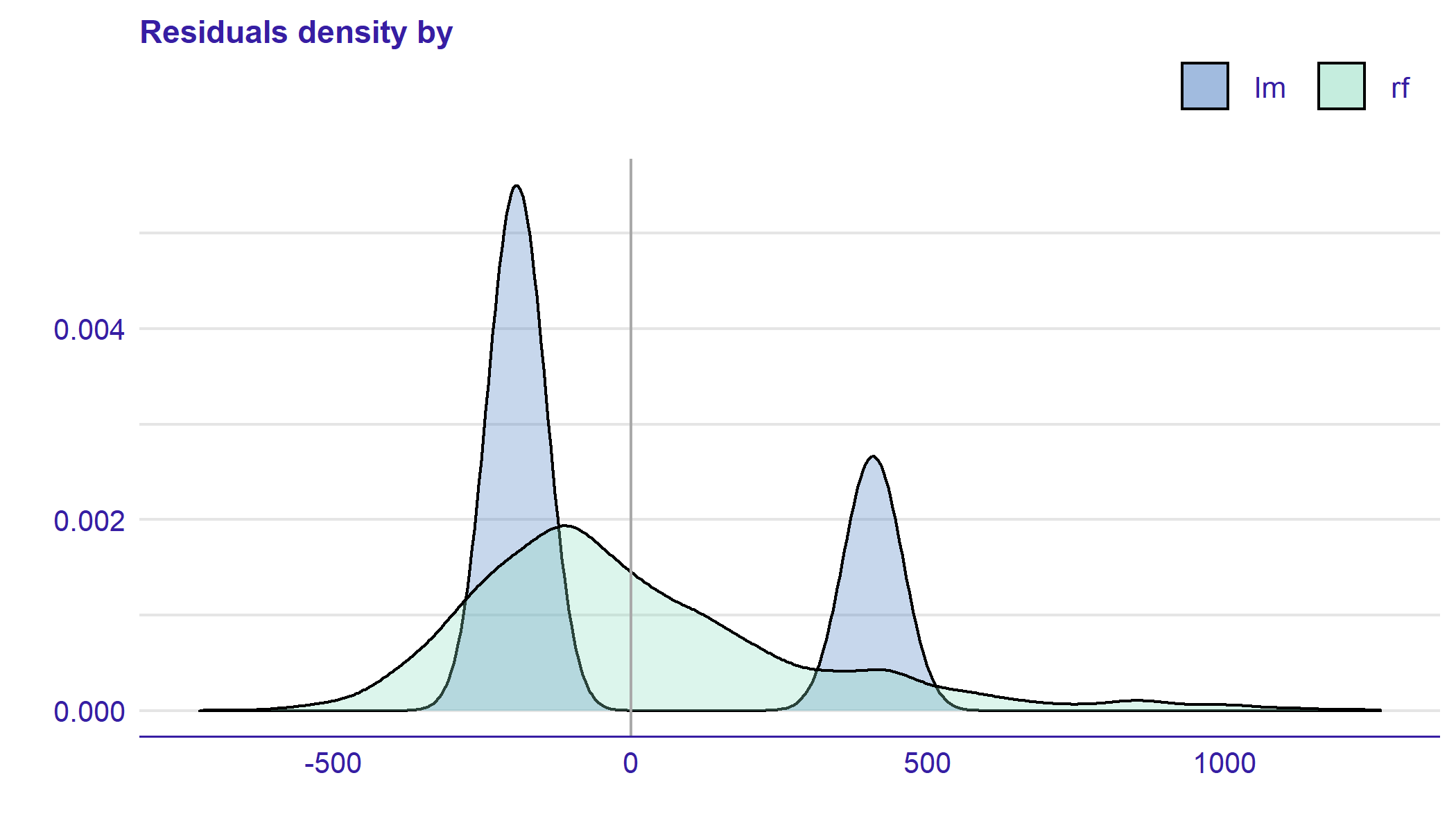}
      \caption[Residual Density Plot]{Residual Density Plot. The density of residuals for the linear model (blue) forms two peaks. There are no residuals with values around zero. Residuals do not follow the normal distribution, which is one of the assumptions of the simple linear regression. There is an~asymmetry of residuals generated by random forest~(green).}
      \label{figure:plotResidualDensity}
\end{figure}

Residual Density plot detects the incorrect behavior of residuals.  Example is presented in Figure~\ref{figure:plotResidualDensity}.
On the plot, there are estimated densities of residuals. Their values are displayed as marks along the x axis.

For some models, the expected shape of density derives from the model assumptions. For example, simple linear model residuals should be normally distributed. However, even if the~model does not have an assumption about the distribution of residuals, such a plot may be informative. If most of the residuals are not concentrated around zero, it is likely that the model predictions~are~biased.

This plot is generated by \code{plot\_residual\_density} function or by \code{plot} function with parameter \code{type = "residual\_density"}. 
\begin{example}
  plot_residual_density(lm_res, rf_res, show_rugs = FALSE)
  # alternative
  plot(lm_res, rf_res, type = "residual_density", show_rugs = FALSE)
\end{example}

%-----------------------------------
\subsubsection{Half-Normal Plot} \label{subsection:half-normal}

The Half-Normal Plot is one of the tools designed to evaluate the goodness-of-fit of a statistical model. It is a~graphical method for comparing two probability distributions by plotting their quantiles against each other.

Points correspond to ordered absolute values of model diagnostic (i.e. standardized residuals) plotted against theoretical order statistics from a half-normal distribution.
Lines mark the~simulated envelope.
The simulation process consists of simulating response variables using the~model matrix, error distribution and parameters of an original model. Then the same model is fitted to data with new simulated response variable. The envelopes are formed on the~basis of the 2.5 and 97.5 percentiles of the~extracted simulated residuals at each value of the~expected order statistic.
Further details on the~half-normal plots with simulated envelopes may be found in \citet{JSSv081i10}.

If residuals come from the normal distribution, they are close to a straight line. However, even if there is no certain assumption of a specific distribution, points still show a certain trend. Simulated envelopes help to verify the correctness of this trend. 
For a good-fitted model, diagnostic values should lay within the envelope.

This plot is generated by \code{plot} function with parameter \code{type = "halfnormal"} or by function  \code{plot\_halfnormal}. The parameter \code{quant = TRUE} transforms y axis into quantile scale.

\begin{example}
  lm_hn <- model_halfnormal(lm_audit)
  
  plot_halfnormal(lm_hn)
  # alternative
  plot(lm_hn, type = "halfnormal")
\end{example}
  
\begin{figure}[H]
    \centering
      \includegraphics[width=0.75\textwidth]{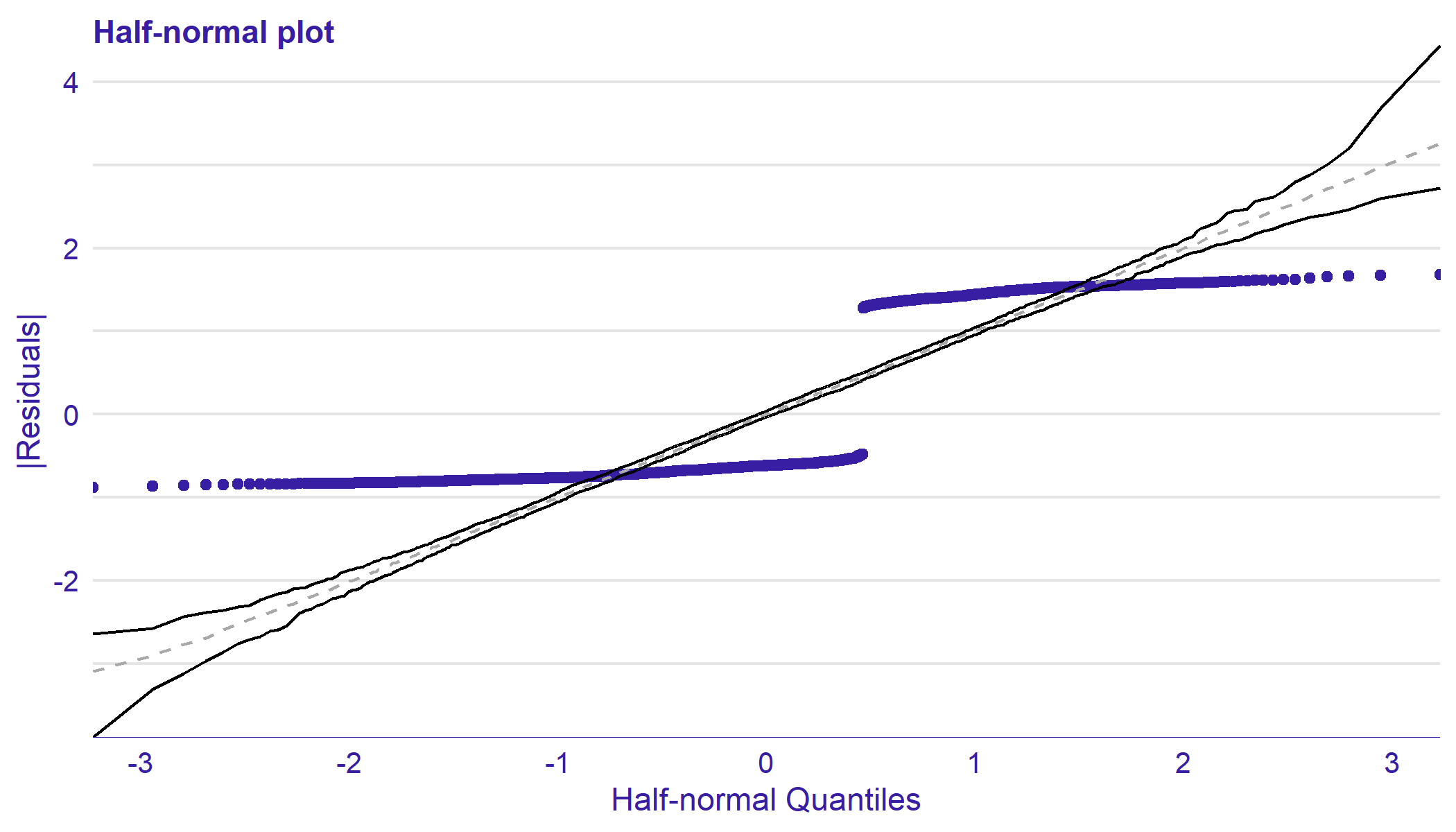}
      \caption[Half-Normal Plot]{Half-Normal plot with simulated envelope. Most of the residuals lie outside the~envelope. They do not behave as we would expect from the simulations. That suggests a poor model fit.}
      \label{figure:plotHalfNormal}
\end{figure}

A useful tool to compare goodness-of-fit of two models is a Half-Normal Score. 

Let us consider $n$ observations and $m$ simulations for each observation. 
We use the following notation: $r_i$ is a residual of $i$-th observation,  $r_{i}^{sim_j}$ is the residual of $j$-th simulation for $i$-th observation.
\begin{definition}
The score of $i$-th observation is given by
\begin{equation}
S_{i} = \sum_{j=1}^m{ \mathbb{1} \{r_{i}^{sim_j} \geq r_i \}} .
\end{equation}
$S_i$ is a number of simulated residuals for observation $i$ that are greater or equal than $r_i$.  
\end{definition}

Value of $S_i$ close to $0$ or $m$ means that the original residual stands out from the simulated ones. 
The closer it is to $\frac{m}{2}$, the less it deviates from the simulated results.

\begin{definition}
  We define Half-Normal score as
  \begin{equation}
    HN = \sum_{i=1}^n |R_i - \frac{m}{2}|.    
  \end{equation}
  $HN$ is the sum of the deviations of $S_i$ from $\frac{m}{2}$.
\end{definition}

Let us note that $HN \in [0, \frac{nm}{2}]$ and lower value of score means better model fit.

HalfNormal score is calculated by \code{score} function with parameter \code{type = "halfnormal"} or by function \code{score\_halfnormal}.
Scores are calculated on the basis of simulated data, so they may differ between function calls.

\begin{example}
  score_halfnormal(lm_audit)
\end{example}

%-----------------------------------
%-----------------------------------
\subsection{Aspect: How similar models are?} \label{model similarity}
In this subsection, we present methods for assessing the similarity of models, understood as the closeness of the residuals structures. The following methods may be used to identify such a similarity or its absence.

%-----------------------------------
\subsubsection{Model PCA Plot}

\begin{figure}[H]
  \centering
      \includegraphics[width=0.8\textwidth]{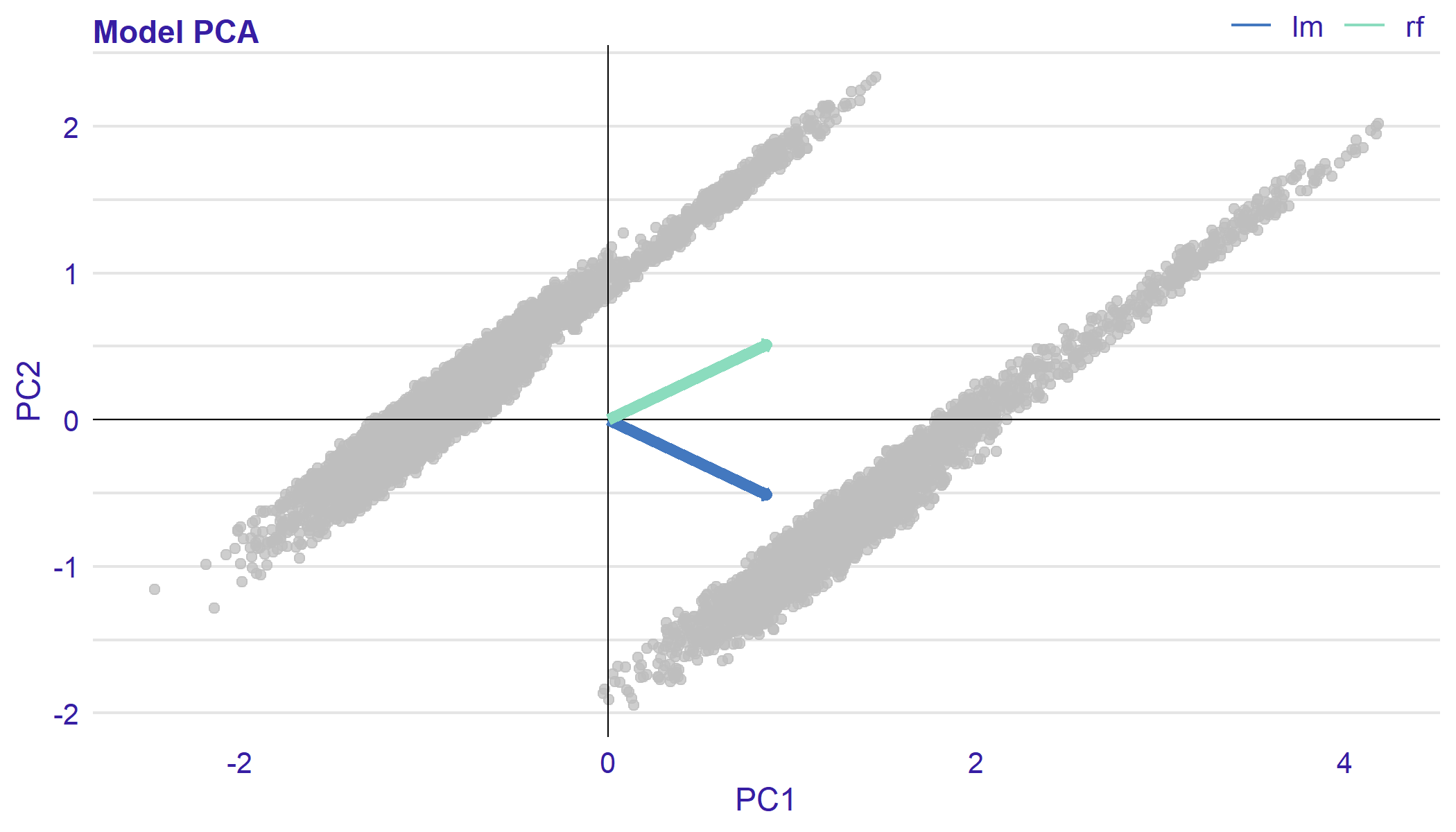}
      \caption[Model PCA Plot]{Model PCA plot. The diagram shows the arrangement of the residuals in two groups. The arrow corresponding to the linear model is almost perpendicular to the groups direction. Therefore, such a division can also be seen in the structure of the residuals of this model. The~arrow corresponding to the random forest is almost parallel to the direction of the residuals. Probably the division into groups does not appear in the residuals of this model.}
      \label{figure:plotModelPCA}
\end{figure}

Model PCA plot can be used to assess the similarity of the models in terms of residuals. 
Example plot is presented in Figure~\ref{figure:plotModelPCA}.

The idea of PCA is reducing the dimension of the data set matrix by creating a set of linearly uncorrelated variables called principal components. At the same time, keeping as much variance as possible \citet{Jolliffe:1986}.

Model PCA plot is a biplot introduced by \citet{doi:10.1093/biomet/58.3.453}. On axis of the plot, there are first two principal components. Grey dots represent observations. Arrows are pointing in the direction of the models projected into a two-dimensional space.
The interposition of arrows provides information about the similarity of models in terms of residuals. 
If they are close to each other, it indicates similar residuals structures.

This plot is generated by \code{plot} function with parameter \code{type = "pca"} or by function  \code{plot\_pca}. 

\begin{example}
  plot_pca(lm_res, rf_res)
  #alternative
   plot(lm_res, rf_res, type ="pca")
\end{example}

\subsubsection{Model Correlation Plot}

\begin{figure}[H]
      \includegraphics[width=\textwidth]{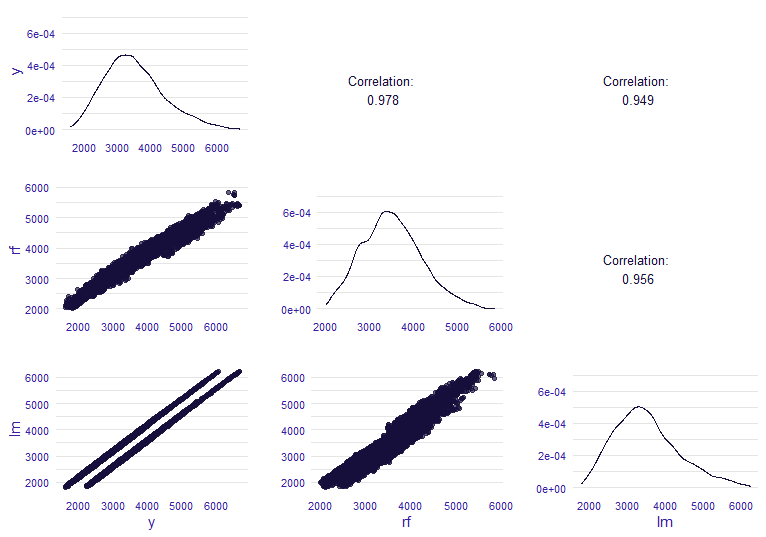}
      \caption[Model Correlation Plot]{Model Correlation plot. The values predicted by random forest and linear model are correlated with each other and with the values observed. The plot does not show any serious problems.}
      \label{figure:plotModelCorrelation}
\end{figure}

Model Correlation plot, presented in Figure~\ref{figure:plotModelCorrelation}, is a pairs plot of observed response and fitted values of different models. Pairs plot first appeared in \citet{doi:10.1080/00949657508810123}. In the diagonal, there are estimated densities, in the upper triangle, the correlations between the different models and between models and observed response. In the lower triangle, there are scatter plots of these values.

This plot is generated by \code{plot} function with parameter \code{type = "correlation"} or by function  \code{plot\_correlation}. 

\begin{example}
  plot_correlation(rf_res, lm_res)
  # alternative
  plot(rf_res, lm_res, type = "correlation")
\end{example}

%-----------------------------------
%-----------------------------------
\subsection{Aspect: Which model has better performance?} \label{model performance}

We consider the model performance more widely than the value of one measure. The values and structures of the residuals also help to assess the quality of the model. In this subsection, we present visual methods of evaluating models performance.

%-----------------------------------
\subsubsection{Predicted Response Plot}

\begin{figure}[H]
\centering
      \includegraphics[width=0.75\textwidth]{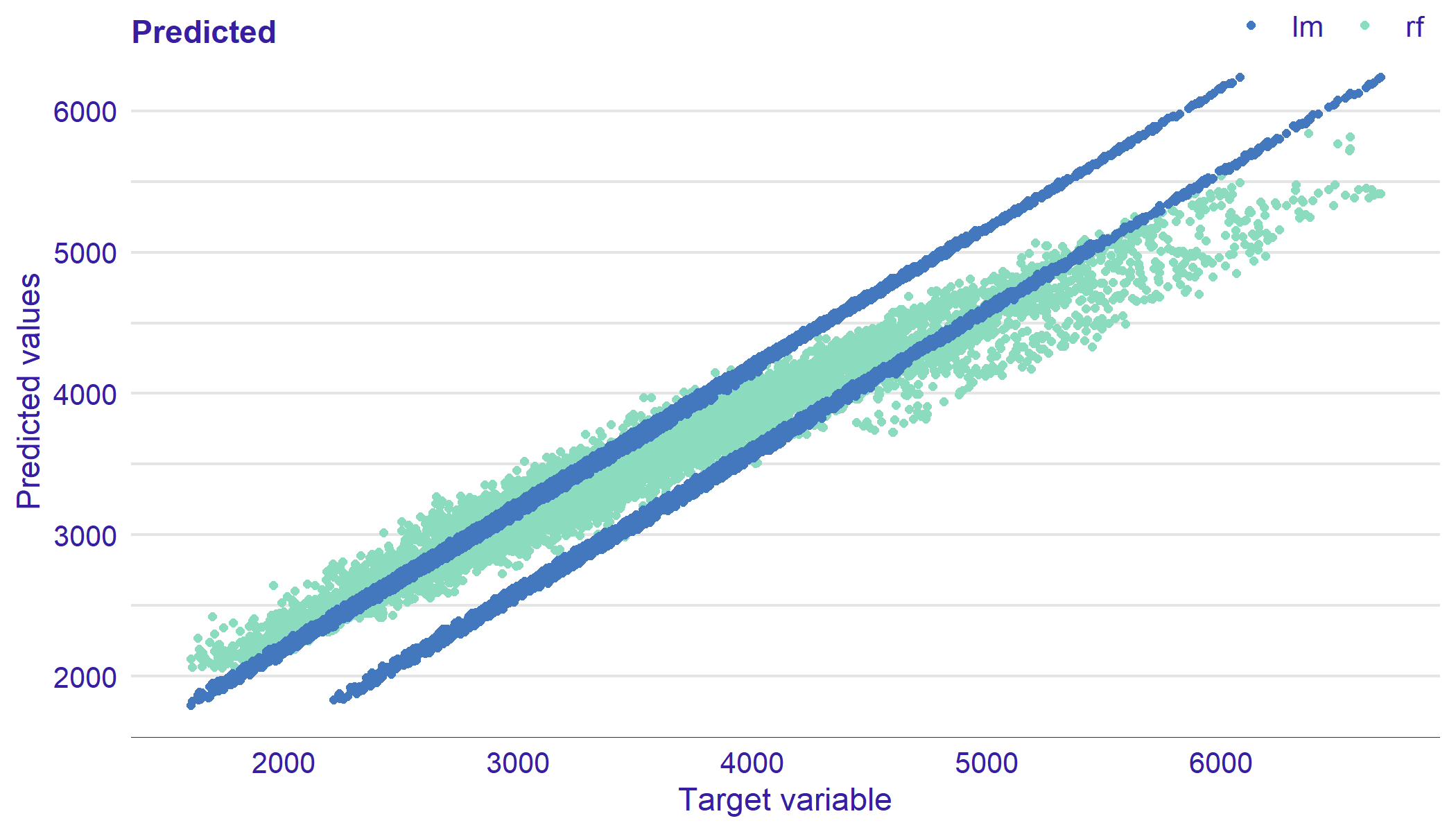}
      \caption[Predicted Response Plot]{redicted Response Plot. The patterns for both models are non-random around the diagonal line. The points corresponding to a random forest (green) show the tendency to underprediction for large values of observed response and over-prediction for small values. Points for linear model (blue) are arranged into two separate groups. That suggests problems with the structures of the models.}
      \label{figure:plotPrediction}
\end{figure}

Predicted Response Plot gives an additional information about model performance. 
It is a~scatter plot of predicted values $\hat{y_i}$ on the y axis against observed response $y_i$ on the x axis. Example plot is presented in Figure~\ref{figure:plotPrediction}. On alterations of this plot on the x-axis are values of one the model variables.
Ideally, all points are close to a diagonal line. For the appropriate model, points should not show any functional dependency. 

This plot is generated by \code{plot} function with parameter \code{type = "prediction"} or by \code{plot\_prediction} function. Other variable on the x axis may be specified by \code{variable} parameter.
\begin{example}
  plot_prediction(lm_res, rf_res)
  # alternative
  plot_prediction(lm_res, rf_res)
\end{example}

%-----------------------------------
\subsubsection{Plots: ROC Curve and LIFT Chart}\label{subsection:ROC}
In this subsection, we present methods for assessing model performance for classification problem. Most common methods for regression models were inspired by tools for classification.

Let us consider a binary classification problem, in which the responses are labeled either as positive ($P$) or negative ($N$). 
True Positive ($TP$) is a case that the prediction is $P$ and the actual value is also $P$.
If the prediction is $P$ and the actual value is $N$ then it is a False Positive ($FP$). 

First, we fit two binary classification models on the \code{PimaIndianDiabetes} data set from the~\code{mlbench} package.
\begin{example}
  library("mlbench")
  data("PimaIndiansDiabetes")
  pima <- PimaIndiansDiabetes
  pima$diabetes <- ifelse(pima$diabetes == "pos", 1, 0)
  model_glm_class <- glm(diabetes~., data = pima, family = binomial)
  library("e1071")
  model_svm_class <- svm(diabetes~., data = pima, probability = TRUE,type = "C-classification")
\end{example}

Next, we create corresponding \code{"model\_audit"} objects. We use created objects to generate example plots in this subsection. 
\begin{example}
  au_glm_class <- audit(model_glm_class, data = pima, y = pima$diabetes)
  
  custom_predict <- function(model, data){
    pred <- predict(model_svm_class, pima, probability=TRUE)
    attr(pred, "probabilities")[,1]
  }

  au_svm_class <- audit(model_svm_class, 
                      data = pima, 
                      y = pima$diabetes,
                      label = "svm", 
                      predict_function = custom_predict)
\end{example}

\begin{figure}[H]
     \centering
      \includegraphics[width=0.5\textwidth]{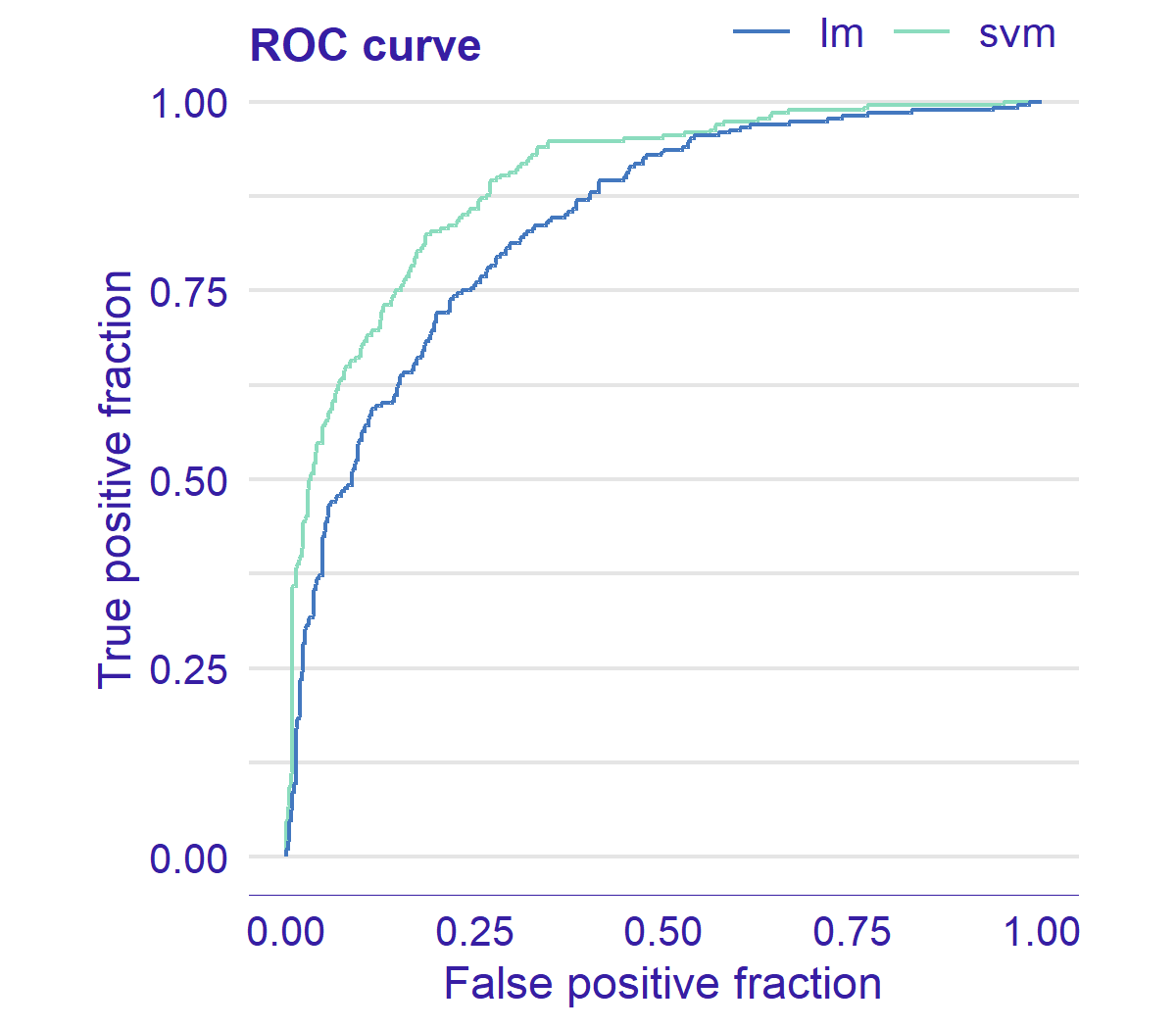}
      \caption[ROC Curve]{Receiver Operating Characteristic curve. The svm (green) curve is mostly above glm (blue). In the remaining part they overlap. This indicates that the support vector machines performs better than linear model.}
      \label{figure:plotROC}
\end{figure}

\strong{Receiver Operating Characteristic (ROC)} curve (See Figure~\ref{figure:plotROC}) is a way of visualising a classifier's performance \citep{Swets1285}. 
It answers the question of how well the model discriminates between the two classes.
The boundary between classes is determined by a threshold value. ROC illustrates the performance of a classification model at various threshold settings.

Let $t \in [0,1]$ be a threshold. We introduce parametric definition of ROC curve.

\begin{definition}
  The ROC curve is a plot of the True Positive Rate ($TPR$) against the False Positive Rate ($FPR$) on a threshold $t$. 
\begin{equation}
    y = TPR(t) = \frac{TP(t)}{\textrm{P(t)}}
\end{equation}
and
\begin{equation}
    x = FPR(t) = \frac{FP(t)}{N(t)}.
\end{equation}
Each point on the ROC curve represents values of $TPR$ and $FPR$ of different thresholds.
\end{definition}

The diagonal line y = x corresponds to a classifier that randomly guess the positive class half the time. Any model that appears in the lower right part of plot performs worse than random guessing. The closer the curve is to the the left border and top border of plot, the more accurate the classifier is.

This plot is generated by \code{plot} function with parameter \code{type = "roc"} or by function  \code{plot\_roc}. 
\begin{example}
  glm_class_me <- model_evaluation(au_glm_class)
  svm_class_me <- model_evaluation(au_svm_class)

  plot_roc(glm_class_me, svm_class_me)
  # altrnative
  plot(glm_class_me, svm_class_me, type = "roc")
\end{example}

ROC curves usually intersect. Therefore, it is not possible to choose a better model explicitly. A good tool to help assess the quality of a classifier is the Area Under the Curve (AUC) \citep{BRADLEY19971145}. The AUC is a part of the area of the unit square. Therefore, its value is always be between $0$ and $1$. The AUC of a classifier is equal to the probability that the classifier will rank a randomly chosen positive instance higher than a randomly chosen negative one \citep{Fawcett:2006:IRA:1159473.1159475}.

AUC may be obtained by \code{score} function with \code{type = "auc"} or \code{score\_auc} function.

\begin{example}
  score_auc(au_glm_class)
  score_auc(au_svm_class)
  # alternative
  score(au_glm_class, type = "auc")
  score(au_svm_class, type = "auc")
\end{example}

%-----------------------------------
\strong{LIFT charts} \citep{Witten:2011:DMP:1972514} also evaluate performance of classification models.

\citet{vuk06roc} introduced parametric definition of LIFT chart. Let $t \in [0,1]$ be~a~threshold.
\begin{definition}
  The LIFT chart is Rate of Positive Prediction ($RPP$) plotted against True Positive (TP) on a threshold $t$.
  \begin{equation}
    y = TP(t)  
  \end{equation}
  and
  \begin{equation}
    x = RPP(t) = \frac{TP(t)+FP(t)}{P+N} .
  \end{equation}
  Each point on the LIFT chart represents values of $TP$ and $RPP$ of different thresholds.
\end{definition}

\begin{figure}[H]
      \centering
      \includegraphics[width=0.5\textwidth]{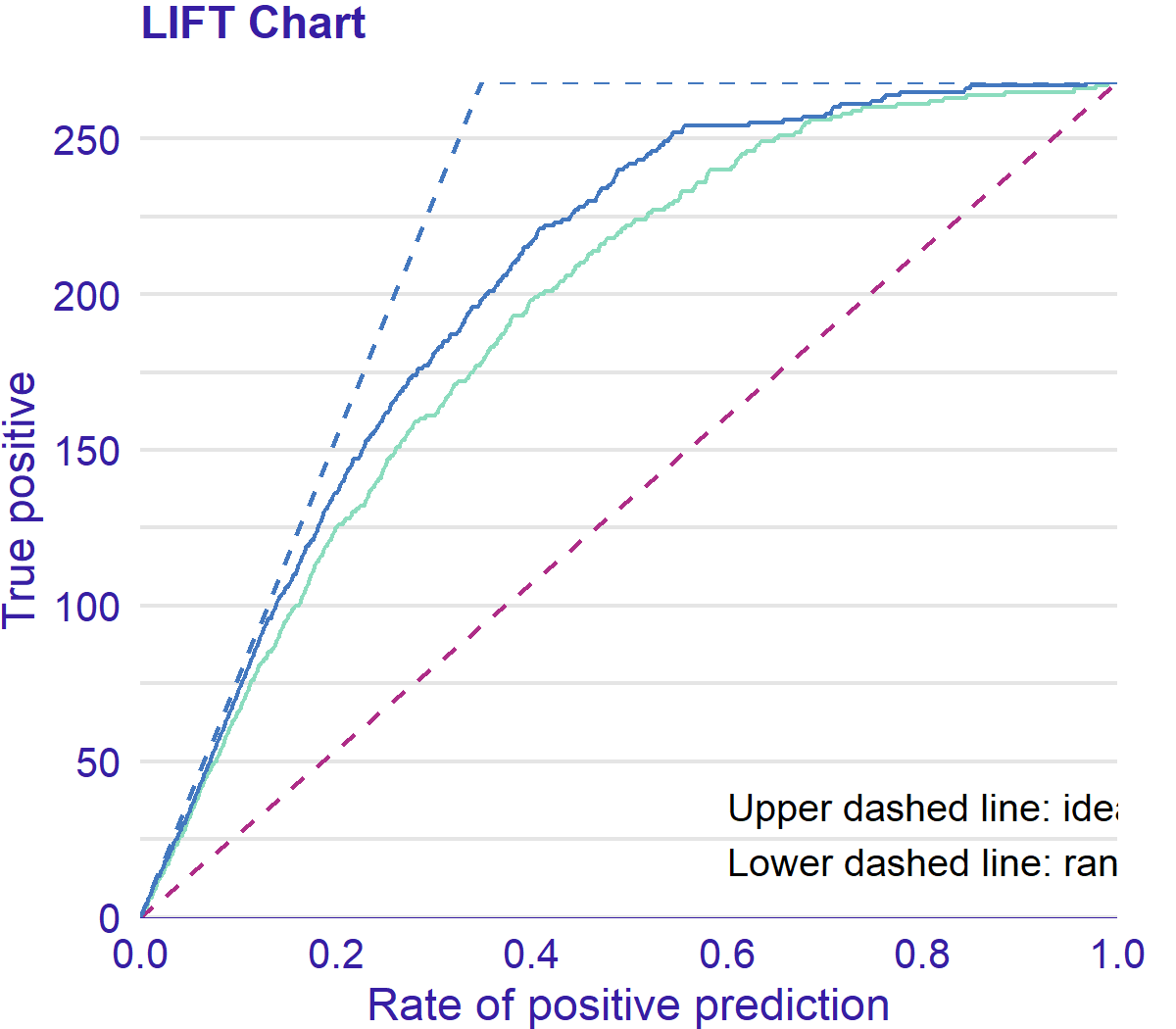}
      \caption[LIFT Chart]{LIFT Chart. Green model (svm) is above the blue (lm), which suggests that it has a~\mbox{better~performance.}}
      \label{figure:plotLIFT}
\end{figure}

As for ROC curve, LIFT (See Figure~\ref{figure:plotLIFT}) illustrates varying the model performance for different thresholds. 
LIFT chart shows how good the classifier distinguishes between two classes.

A random and ideal models are represented by black and orange curves, respectively. The closer the LIFT get to the orange curve, the better a model is.

This plot is generated by \code{plot} function with parameter \code{type = "lift"} or by function~\code{plot\_llift}. 

\begin{example}
  plot(au_glm_class, au_svm_class, type = "lift")
\end{example}

%-----------------------------------
\subsubsection{REC Curve Plot}\label{subsection:REC}

\begin{figure}[H]
\centering
      \includegraphics[width=0.75\textwidth]{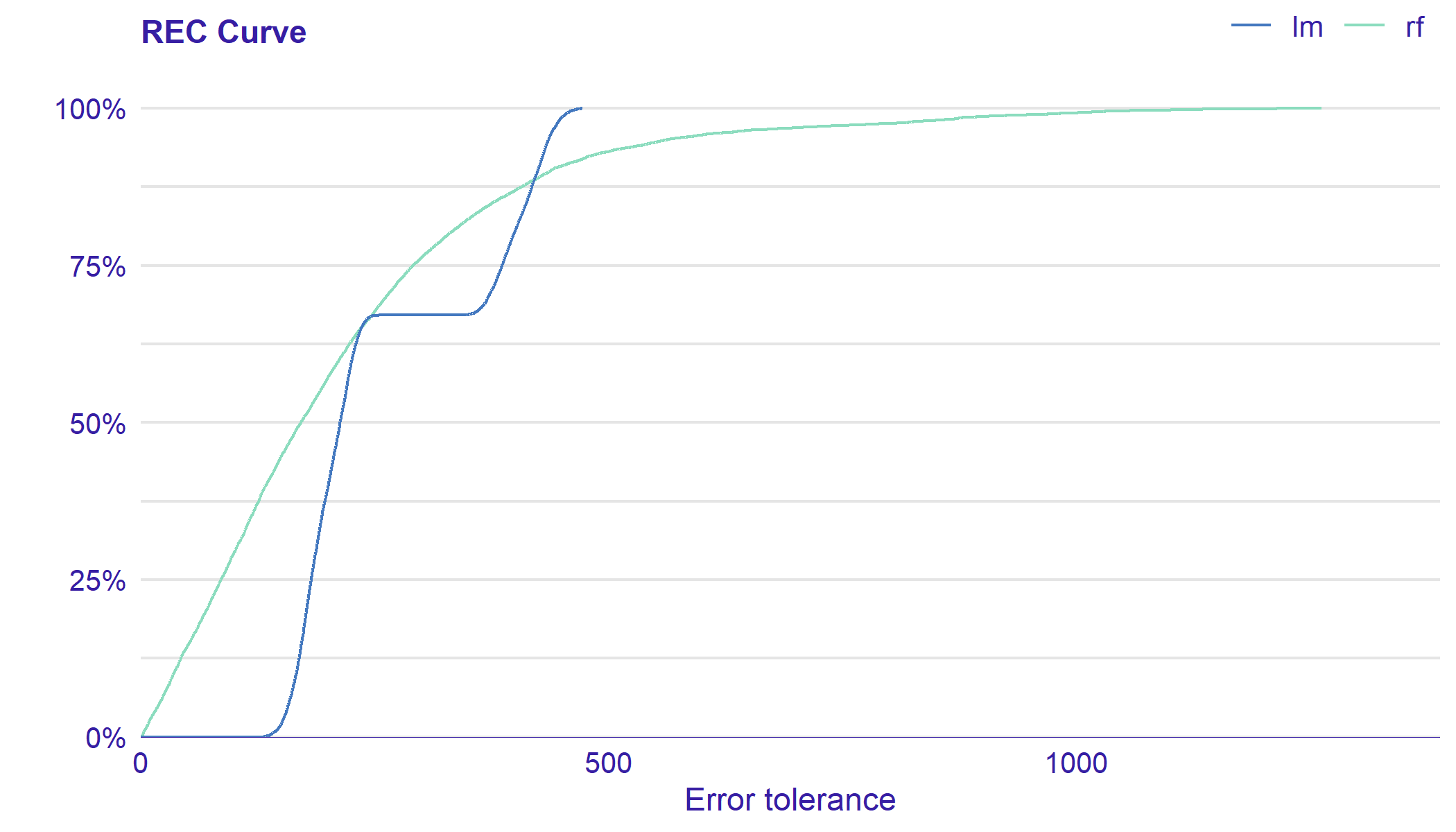}
      \caption[REC Curve]{REC curve. Curve for linear model (blue) suggests that model is biased. It has poor accuracy when the tolerance $\epsilon$ is small. However, once $\epsilon$ exceeds the error tolerance $130$ the~accuracy rapidly increases. The random forest (green) has a stable increase of accuracy when compared to the linear model. However, there is s fraction of large residuals.}
      \label{figure:plotREC}
\end{figure}

Regression Error Characteristic (REC) curve (See Figure~\ref{figure:plotREC}) is a generalization of Receiver Operating Characteristic (ROC) curve for binary classification described in subsection~\ref{subsection:ROC}. 

REC curve estimates the Cumulative Distribution Function of the error. On the x axis of the~plot there is an~error tolerance. On the y axis there is an accuracy at the given tolerance level. \citet{Bi2003RegressionEC} define the accuracy at tolerance $\epsilon$ as a percentage of observations predicted within the tolerance $\epsilon$. In other words, residuals larger than $\epsilon$ are considered as~errors.

Let consider pairs $(y_i, x_i)$ as at the beginning of the Chapter~\ref{audit}. \citet{Bi2003RegressionEC} define an accuracy as follows.
\begin{definition}\label{def:REC}
An accuracy at tolerance level $\epsilon$ is given by
\begin{equation}
    acc(\epsilon) = \frac{|\{ (x,y): loss(f(x_i),y_i) \leq \epsilon, i = 1,...,n \}|}{n}.
\end{equation}
\end{definition}
REC Curves implemented in the \pkg{auditor} are plotted for a special case of Definition \ref{def:REC} where the loss is defined as 
  \begin{equation}
     loss(f(x_i),y_i) = |f(x_i) - y_i| = |r_i|.    
  \end{equation}

The shape of the curve illustrates the behavior of errors. The quality of the model can be evaluated and compared for different tolerance levels.  
The stable growth of the accuracy does not indicate any problems with the model. 
A small increase of accuracy near $0$ and areas, where the growth is fast, signalize bias of the model predictions.

This plot is generated by \code{plot} function with parameter \code{type = "rec"} or by \code{plot\_rec} function. 
\begin{example}
  plot_rec(lm_res, rf_res)
  # alternative
  plot_rec(lm_res, rf_res)
\end{example}

As often it is difficult to compare models on the plot, there is an REC score implemented in the \pkg{auditor}.
This score is the Area Over the REC Curve (AOC), which is a biased estimate of the expected error for a~regression model. As \citet{Bi2003RegressionEC} proved, AOC provides a~measure of the overall performance of regression model.

Scores may be obtained by \code{score} function with \code{type = "rec"} or \code{score\_rec} function.
\begin{example}
  score_rec(lm_audit)
  score_rec(rf_audit)
  # alternative 
  score(lm_audit, type = "rec")
  score(rf_audit, type = "rec")

\end{example}

%-----------------------------------
\subsubsection{Two-sided ECDF Plot}

\begin{figure}[H]
\centering
      \includegraphics[width=0.75\textwidth]{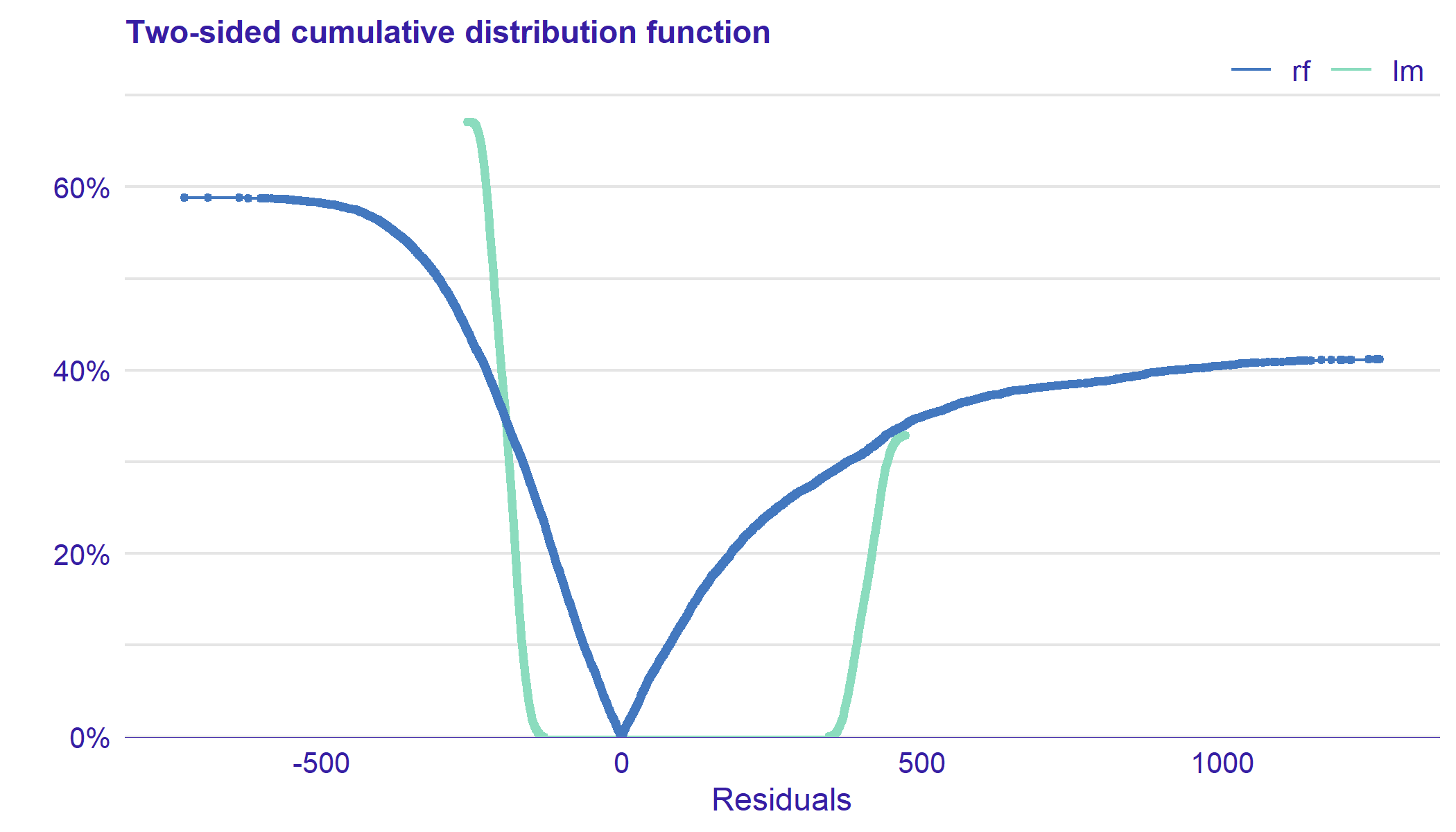}
      \caption[Two-Sided ECDF Plot]{Two-sided ECDF plot. The plot shows that majority of residuals for the random forest (green) is smaller than residuals for the linear model (blue). However, random forest has also fractions of large residuals.}
      \label{figure:plotTwoSidedECDF}
\end{figure}

Two-sided ECDF plot (See Figure~\ref{figure:plotTwoSidedECDF}) shows an Empirical Cumulative Distribution Functions (ECDF) for positive and negative values of residuals separately. 

Let $x_1, ..., x_n$ be a random sample from a cumulative distribution function $F(t)$.
The following definition comes from  \citet{van2000asymptotic}.

\begin{definition}
The empirical cumulative distribution function is given by
  \begin{equation}
    F_n(t) = \frac{1}{n} \sum_{i=1}^n \mathbb{1} \{ x_i \leq t\}.  
  \end{equation}
Empirical cumulative distribution function presents a fraction of observations that are less or equal than $t$. It~is an~estimator for the cumulative distribution function $F(t)$.
\end{definition}

On the positive side of the x-axis, there is the ECDF of positive values of residuals. 
On the~negative side, there is a transformation of ECDF: 
\begin{equation}
  F_{rev}(t) = 1 - F(t).
\end{equation}
Let $n_N$ and $n_P$ be numbers of negative and positive values of residuals, respectively.
Negative part of the plot is normalized by multiplying it by the ratio of the $n_N$  over $n_N + n_P$. 
Similarly, positive part is normalized by multiplying it by the ratio of the $n_P$  over $n_N + n_P$.
Due to the scaling, the ends of the curves add up to $100\%$ in total. 

The plot shows the distribution of residuals divided into groups with  positive and negative values. It helps to identify the asymmetry of the residuals. Points represent residuals.

This plot is generated by \code{plot\_tsecdf} function or by \code{plot} function with parameter \code{type = "tsecdf"}. 

\begin{example}
  plot_tsecdf(rf_res, lm_res)
  # alternative
  plot(rf_res, lm_res, type = "tsecdf")
\end{example}

%-----------------------------------
\subsubsection{RROC Curve Plot}\label{subsection:RROC}

\begin{figure}[H]
      \includegraphics[width=0.75\textwidth]{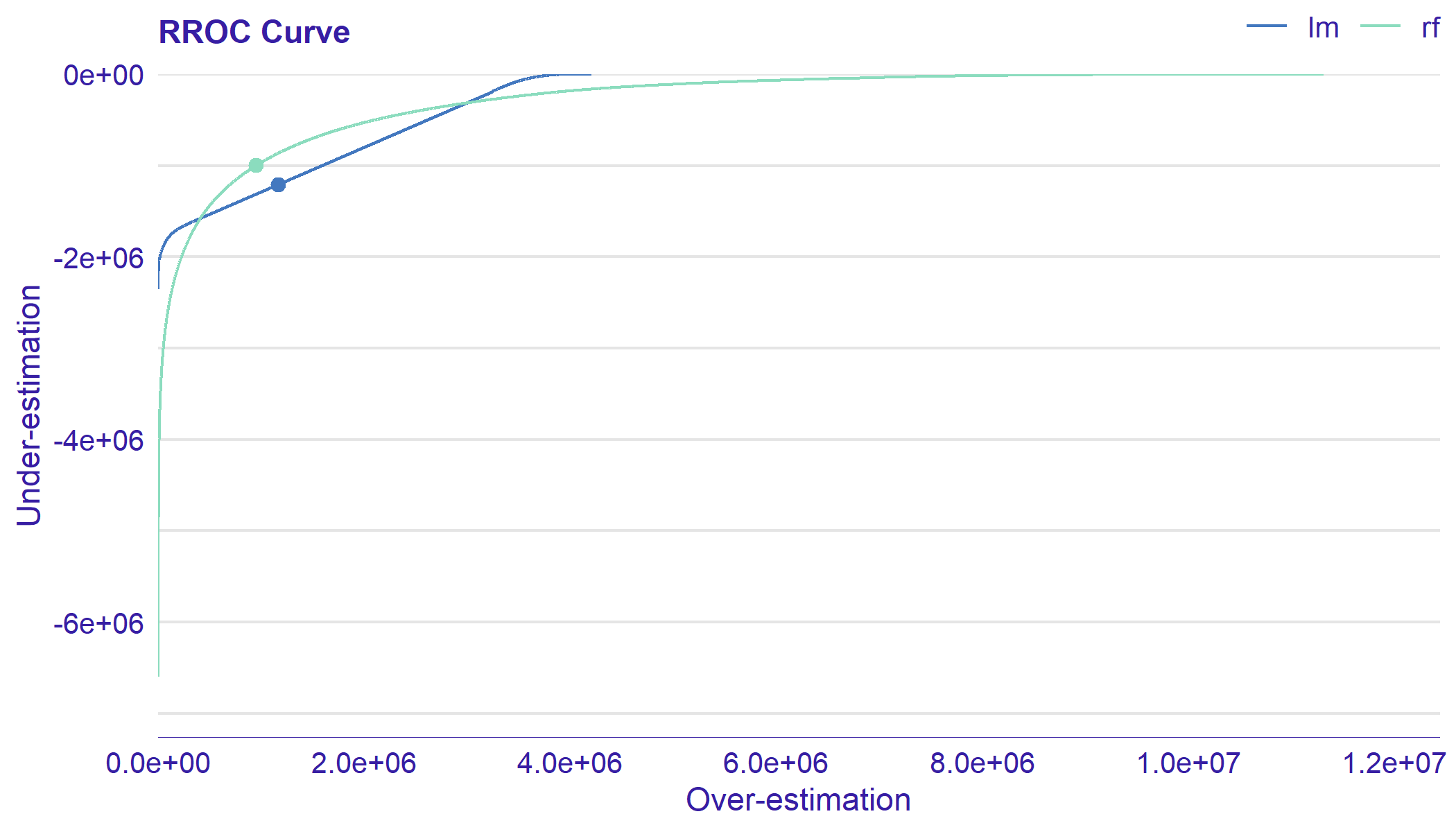}
      \caption[RROC Curve]{The Regression Receiver Operating Characteristic (RROC) curve for regression. For shifts around $0$ random forest (green) performs better than linear regression (blue).}
      \label{figure:plotRROC}
\end{figure}

The Regression Receiver Operating Characteristic (RROC) curve for regression shows model asymmetry, which is an inequality in the number and values of positive and negative residuals. It may be useful for problems, where there is an asymmetric cost of errors. An~example of RROC plot is presented in Figure~\ref{figure:plotRROC}.

The RROC curve was introduced by \citet{HERNANDEZORALLO20133395}.
The base of plot is a~shift~$s$.~It is an equivalent to the threshold for ROC curves presented in subsection~\ref{subsection:ROC}. For each observation we calculate new prediction: $\hat{y}'^s = \hat{y} + s$. 
\citet{HERNANDEZORALLO20133395} define over- and under-estimation as follow.

\begin{definition}
  Over-estimation depending on shift $s$ is given by
  \begin{equation} 
    OVER(s) = \sum_{i=1}^n{r_i \mathbb{1} \{ r_i + s > 0 \} }.
  \end{equation}
\end{definition}
\begin{definition}
  Under-estimation depending on shift $s$ is given by
  \begin{equation} 
    UNDER(s) = \sum_{i=1}^n{r_i \mathbb{1} \{ r_i + s < 0 \} }.
  \end{equation}
\end{definition}
The RROC plot consists of under-estimation on the y axis against over-estimation on the~x~axis. The shift equals $0$ is represented by a dot.

The shape of the curve illustrates the behavior of errors. The quality of the model can be evaluated and compared for different shifts. The AOC measures the model insensitivity to asymmetric costs of errors.

This plot is generated by \code{plot} function with parameter \code{type = "rroc"} or \mbox{by~\code{plot\_rroc}~function.}
\begin{example}
  plot_rroc(lm_res, rf_res)
  # alternative
  plot(lm_res, rf_res, type = "rroc")
\end{example}

As often it is difficult to compare AOC on the plot, there is an RROC score implemented in the auditor package. This score is the Area Over the RROC Curve (AOC). As  \citet{HERNANDEZORALLO20133395} prove AOC equals to the variance of the errors multiplied by a $\frac{n^2}{2}$ which is independent of the model. Lower values for AOC are better.

Score may be obtained by \code{score} function with \code{type = 'rroc'} or \code{score\_rroc} function.
\begin{example}
  score_rroc(lm_audit)
  score_rroc(rf_audit)
  # alternative
  score(lm_audit, type = "rroc")
  score(rf_audit, type = "rroc")
\end{example}

%-----------------------------------

\subsubsection{Model Ranking Plot}

\begin{figure}[H]
\centering
      \includegraphics[width=0.5\textwidth]{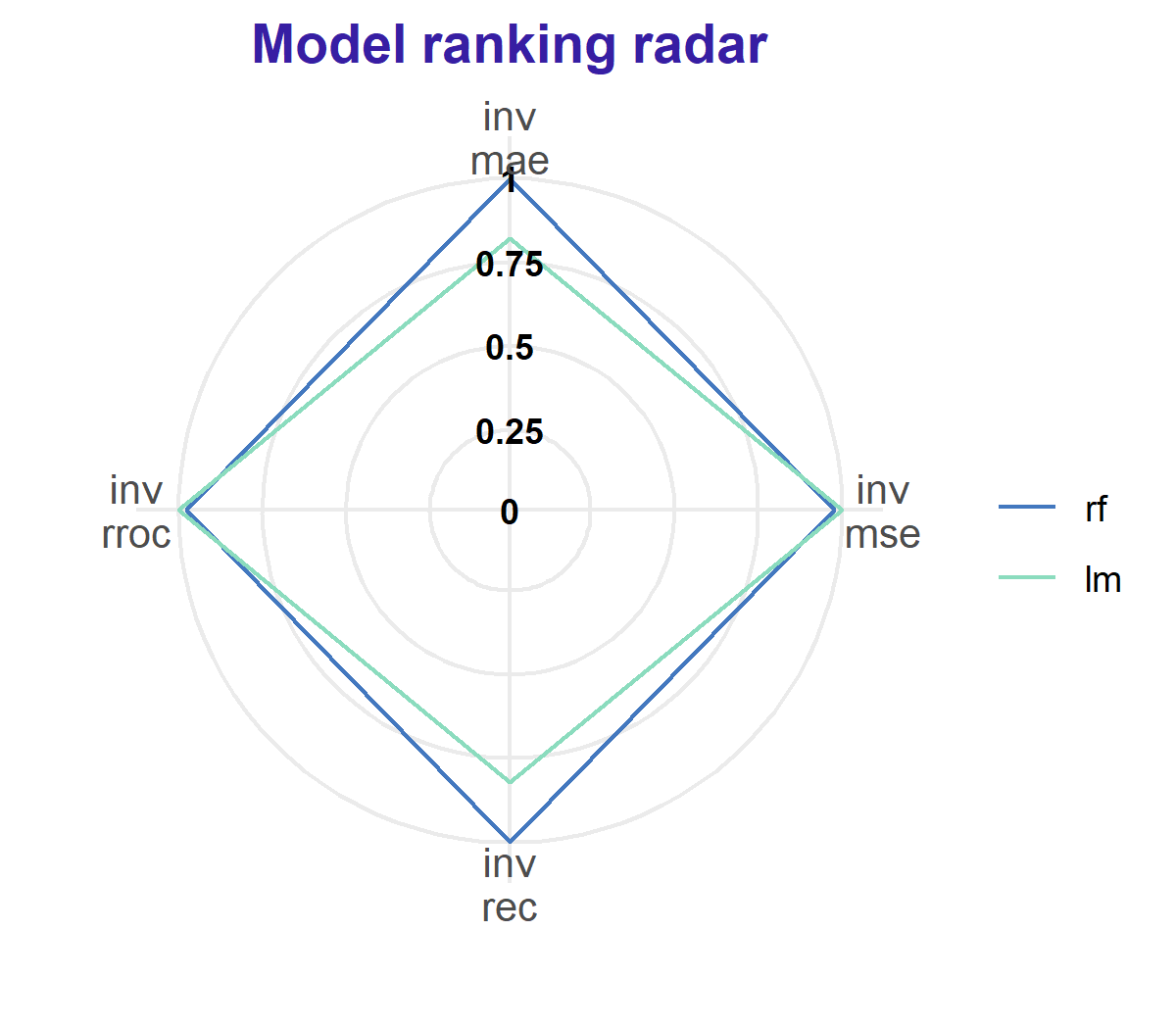}
      \caption[Model Ranking Plot]{Model Ranking Plot. Random forest (green) has better performance in aspect of MAE and REC scores, while linear model (blue) is better in aspect of MSE and RROC scores.}
      \label{figure:plotModelRanking}
\end{figure}

In this subsection, we propose a Model Ranking plot, which compares models performance across multiple measures (see Figure~\ref{figure:plotModelRanking}). The implemented measures coincide with the~scores listed in the Chapter~\ref{architecture}.

In previous subsections, we introduced the following scores: 
\begin{itemize}
    \item DW (subsection~\ref{subsection:autocorrelation}), 
    \item Peak (subsection~\ref{subsection:scalelocation}), 
    \item Half-Normal (subsection~\ref{subsection:half-normal}), 
    \item REC(subsection~\ref{subsection:REC}),
    \item ROC (subsection~\ref{subsection:ROC}), 
    \item RROC(subsection~\ref{subsection:RROC}), 
    \item Runs (subsection~\ref{subsection:autocorrelation}).
\end{itemize}

Additional scores implemented in the auditor are: 

\begin{itemize}
    \item MAE (Mean Absolute Error)
            \begin{equation}
                 MAE = \frac{1}{n} \sum_{i=1}^n \mid r_i \mid ,
            \end{equation}

    \item MSE (Mean Squared Error)
        \begin{equation}
            MSE = \frac{1}{n}\sum_{i=1}^n r_i^2 ,
        \end{equation}

    \item RMSE (Root Mean Squared Error)
        \begin{equation}
            RMSE = \sqrt{\frac{1}{n}\sum_{i=1}^n r_i^2} .
        \end{equation}

\end{itemize}

Model Ranking Radar plot consists of two parts. On the left side, there is a radar plot. Colors correspond to models, edges to values of scores. Score values are inverted and rescaled~to~$[0,1]$.

Let us use the following notation:  $ m_i \in \mathcal{M}$ is a model in a finite set of models $\mathcal{M}$, where $|\mathcal{M}| = k$, 
$score: \mathcal{M} \to \mathbb{R}$ is a scoring function for the model under consideration, that higher value means worse model performance.
The $score(m_i)$ is a performance of model $m_i$. 
\begin{definition}
  We define the inverted score of model $m_i$ as
  \begin{equation}\label{invscore}
    invscore(m_i) = \frac{1}{score(m_i)} \min_{j=1...k}{score(m_j)}.
  \end{equation}
\end{definition}
Models with the larger $invscore$ are closer to the centre. Therefore, the best model is located the farthest from the center of the plot.

On the right side of the plot there is a table with results of scoring. In the third column, there are scores scaled to one of the models. 

Let $m_l \in \mathcal{M}$ where $l \in \{ 1,2, ..., k \}$ be a model to which we scale. 
\begin{definition}
  We define the scaled score of model $m_i$ to model $m_l$ as 
  \begin{equation}
    scaled_l(m_i) = \frac{score(m_l)}{score(m_i)}.     
  \end{equation}
\end{definition}
As values of $scaled_l(m_l)$ are always between $0$ and $1$, comparison of models is easy, regardless of the ranges of scores.
 
This plot is generated by \code{plot} function with parameter \code{type = "radar"} or by function  \code{plot\_radar}. 
The scores included in the plot may be specified by \code{scores} parameter.

\begin{example}
  lm_mp <- model_performance(lm_audit)
  rf_mp <- model_performance(rf_audit)
  
  plot_radar(lm_mp, rf_mp)
  # alternatve
  plot(lm_mp, rf_mp, type = "radar")
\end{example}
 
\begin{figure}[H]
\centering
      \includegraphics[width=0.5\textwidth]{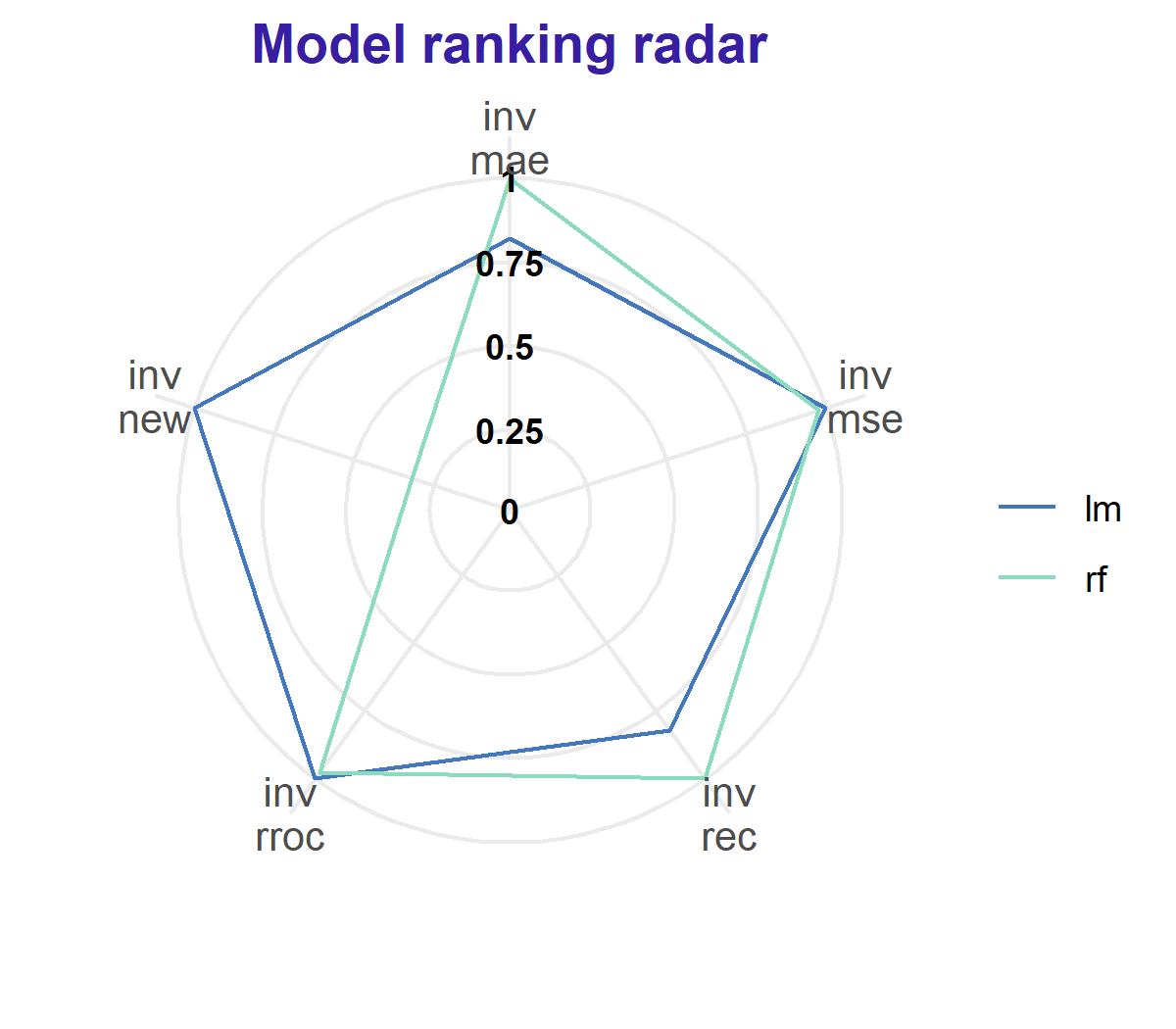}
      \caption[Extended Model Ranking Plot]{Radar plot with model scores. In terms of new score, the random forest (green) has better performance than the linear model (blue).}
      \label{figure:plotModelRanking2}
\end{figure}

There is also a possibility to add custom scores (Figure~\ref{figure:plotModelRanking2}). They may be provided by parameter \code{new.score}. It requires a named list of functions that take one argument: object of class \code{"model\_audit"} and return a~numeric value. The measure calculated by the function should have the property that lower score value indicates better model.

\begin{example}
  new_score <- function(object){sum((object$y - object$y_hat)^4)}

  rf_mp <- model_performance(rf_audit, new_score = list("new" = new_score))
  lm_mp <- model_performance(lm_audit, new_score = list("new" = new_score))

  plot_radar(rf_mp, lm_mp)
  # alternative
  plot(rf_mp, lm_mp, type = "radar")

\end{example}

%-----------------------------------
%-----------------------------------

\subsection{Aspect: Influence of Observations} \label{oher questions}

In this subsection, we focus on the impact of individual observation on a model.

%-----------------------------------
\subsubsection{Cook's Distances Plot}\label{subsection:Cooks}

\begin{figure}[H]
      \includegraphics[width=0.75\textwidth]{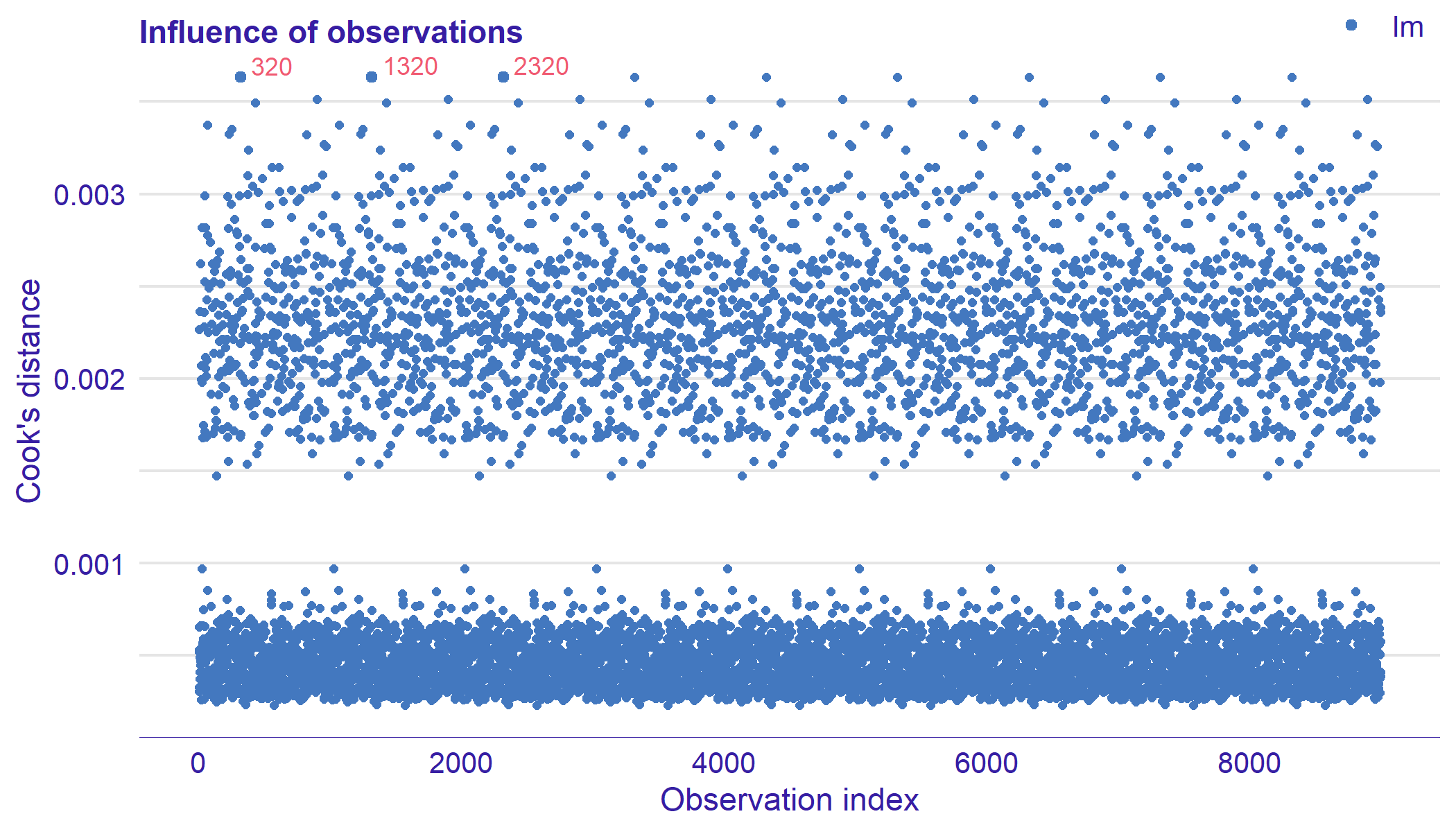}
      \caption[Cook's Distances Plot]{Cook's Distances plot. The 3 observations with the highest values of the Cook's distance are marked. However, they do not significantly differ from others.}
      \label{figure:plotCooksDistance}
\end{figure}

Cook's distances plot presented in Figure~\ref{figure:plotCooksDistance} is a tool for identifying observations that may negatively affect the model. They can be also used for indicating regions of the design space where it would be good to obtain more observations. Data points indicated by Cook’s distances are worth checking for validity.

Let us extend the notation provided iat the beginning of the Chapter~\ref{audit}. We denote the number of predictors as $p$ and the mean squared error as $s^2$. We consider recalculated model that is fitted on the original data set with removed $j$-th observation. Let $\hat{y}_{i(j)}$ be the prediction of such model calculated for $i$-th observation. 
\citet{10.2307/1268249} defined the influence of a single observation as follows.
\begin{definition}
Cook's distance of $i$-th observation is
  \begin{equation}
    D_i = \frac{\sum_{j=1} (\hat{y_j} - \hat{y}_{i(j)})^2 }{ p s^2 }. 
  \end{equation}
Cook’s Distances are calculated by removing the $i$-th observation from the data and recalculating the model. It shows an influence of $i$-th observation on the model.
\end{definition}

\citet{10.2307/1268249} proved that for linear models Cook's distance may be computed in an alternative and computationally convenient way. Let $\mathbf{X}$ be the design matrix and $\mathbf{H = X(X'X)^{-1}X'}$ be a projection matrix. The leverage $h_i$ is the $i$-th diagonal element of $\mathbf{H}$.
$D_i$ is obtained by
\begin{equation}
    D_i = \frac{r_i^2}{p s^2} \frac{h_i}{(1-h_i)^2}.
\end{equation}
    
This plot is generated by \code{plot} function with parameter \code{type = "cooksdistance"} or by function  \code{plot\_cooksdistance}. For model of class \code{"lm"} and \code{"glm"} the distances are computed from the diagonal elements of the hat matrix. For other models they are computed directly from~the~definition.

\begin{example}
  lm_oi <- model_cooksdistance(lm_audit)

  plot_cooksdistance(lm_oi)
  # alternative
  plot(lm_oi, type = "cooksdistance")
\end{example}

%% file: chapters/use_case.tex
\section{Example of model audit} \label{use-case}

This section contains a use-case of model audit. We show an example, where the choice of a model on the basis of a single measurement is not obvious. We illustrate, how to use \pkg{auditor} to analyze and compare models. In addition, we present how to use residual analysis to identify outliers and better understand a~model structure.

\subsection{The artificial data set }
In order to present the \pkg{auditor} we created an artificial data set \code{auditorData}. It consists of 2000 observations. We generated first 1998 as follows.
Let 
\begin{itemize}
    \item  $X_1 = (x_1^{(1)}, x_2^{(1)}, ..., x_{1998}^{(1)})$ where $ x_i^{(1)} \sim \mathcal{U}[0, 1] $,
    \item  $X_2 = (x_1^{(2)}, x_2^{(2)}, ..., x_{1998}^{(2)})$ where $ x_i^{(2)} \sim \mathcal{U}[0, 1] $, 
    \item  $X_3 = (x_1^{(3)}, x_2^{(3)}, ..., x_{1998}^{(3)})$ where $ x_i^{(3)} \sim \mathcal{U}[0, 1] $, 
    \item  $X_4 = (x_1^{(4)}, x_2^{(4)}, ..., x_{1998}^{(4)})$ where $ x_i^{(4)} $ is from discrete probability distribution, \\ $P(x_i^{(4)} = 0) = 0.5$, $P(x_i^{(4)} = 1) = 0.35$, $P(x_i^{(4)} = 4) = 0.15$, 
    \item  $\epsilon = (\epsilon_1, \epsilon_2, ..., \epsilon_{1998})$ where $\epsilon_i \sim \mathcal{N}(0, 0.5)$
\end{itemize}
 for $i \in {1, 2, ..., 1998}$.

Let us simulate a response $Y = (y_1, y_2, ..., y_{1998})$ as a function of five arguments: $X_1$, $X_2$, $X_3$, $X_4$, $\epsilon$, where
\begin{equation}
    y_i = 20(x_i^{(1)}-1)^2 + 2(x_i^{(2)}-0.25)(x_i^{(2)}-0.5)(x_i^{(2)}-1) + {22 x_i^{(3)} - 1} + 5 x_i^{(4)} x_i^{(1)} + \epsilon_i .
\end{equation}

The $1999$-th and $2000$-th observations are added manually. They are meant to be outliers. Their values are 
\begin{equation}
  (y_{1999}, x_{1999}^{(1)}, x_{1999}^{(2)}, x_{1999}^{(3)}, x_{1999}^{(4)}) = (92, 0.32, 0.21, 0.1, 0)   
\end{equation}
 and 
\begin{equation}
  (y_{2000}, x_{2000}^{(1)}, x_{2000}^{(2)}, x_{2000}^{(3)}, x_{2000}^{(4)}) = (98, 0.86, 0.82, 0.85, 0),
\end{equation}
respectively. 

First four of simulated variables are treated as continuous while the fifth one is categorical. 
In the Table~\ref{table:auditorData} we present first six observations from the \code{auditorData} included in the \pkg{auditor} package.
\begin{example}
  library("auditor")
  data("auditorData")
  head(auditorData)
\end{example}
\begin{table}[!htbp]
    \begin{center}
        \begin{tabular}{lllll}
            \hline
            y     & X1   & X2   & X3   & X4 \\ \hline
        25.08  & 0.10 & 0.11 & 0.45 & 0 \\
        11.43 & 0.78 & 0.38 & 0.52 & 0 \\
        31.36 & 0.61 & 0.33 & 0.78 & 4 \\
        0.84 & 0.96 & 0.01 & 0.09 & 0 \\
        22.23 & 0.96 & 0.32 & 0.84 & 1 \\
        29.35 & 0.22 & 0.16 & 0.83 & 0 \\ \hline
        \end{tabular}
        \caption{First 6 observations from the data set \code{auditorData}.}
        \label{table:auditorData}
    \end{center}
\end{table}
Our goal is to predict Y based on selected variables X1, X2, X3 and X4.

\subsection{Fitting models}

We fit 3 models and audit them. They are: simple linear regression, random forest, and support vector regression. We use  \code{randomForest} function from \CRANpkg{randomForest} package \citep{randomForest} and \code{svm} function from \CRANpkg{e1071} package \citep{e1071}.

\begin{example}
  model_lm <- lm(y ~ ., auditorData)

  library("randomForest")
  set.seed(1994)
  model_rf <- randomForest(y~., auditorData)

  library("e1071")
  model_svm <- svm(y ~ ., auditorData)
\end{example}  

Next step is creating three \code{"modelAudit"} objects, corresponding to the models.

\begin{example}
 au_lm <- audit(model_lm, data = auditorData, y = auditorData$y)
 au_rf <- audit(model_rf, data = auditorData, y = auditorData$y, label = "rf")
 au_svm <- audit(model_svm, data = auditorData, y = auditorData$y, label = "svm")
\end{example}

\subsection{Model audit}

\begin{figure}[H]
\centering
      \includegraphics[width=0.5\textwidth]{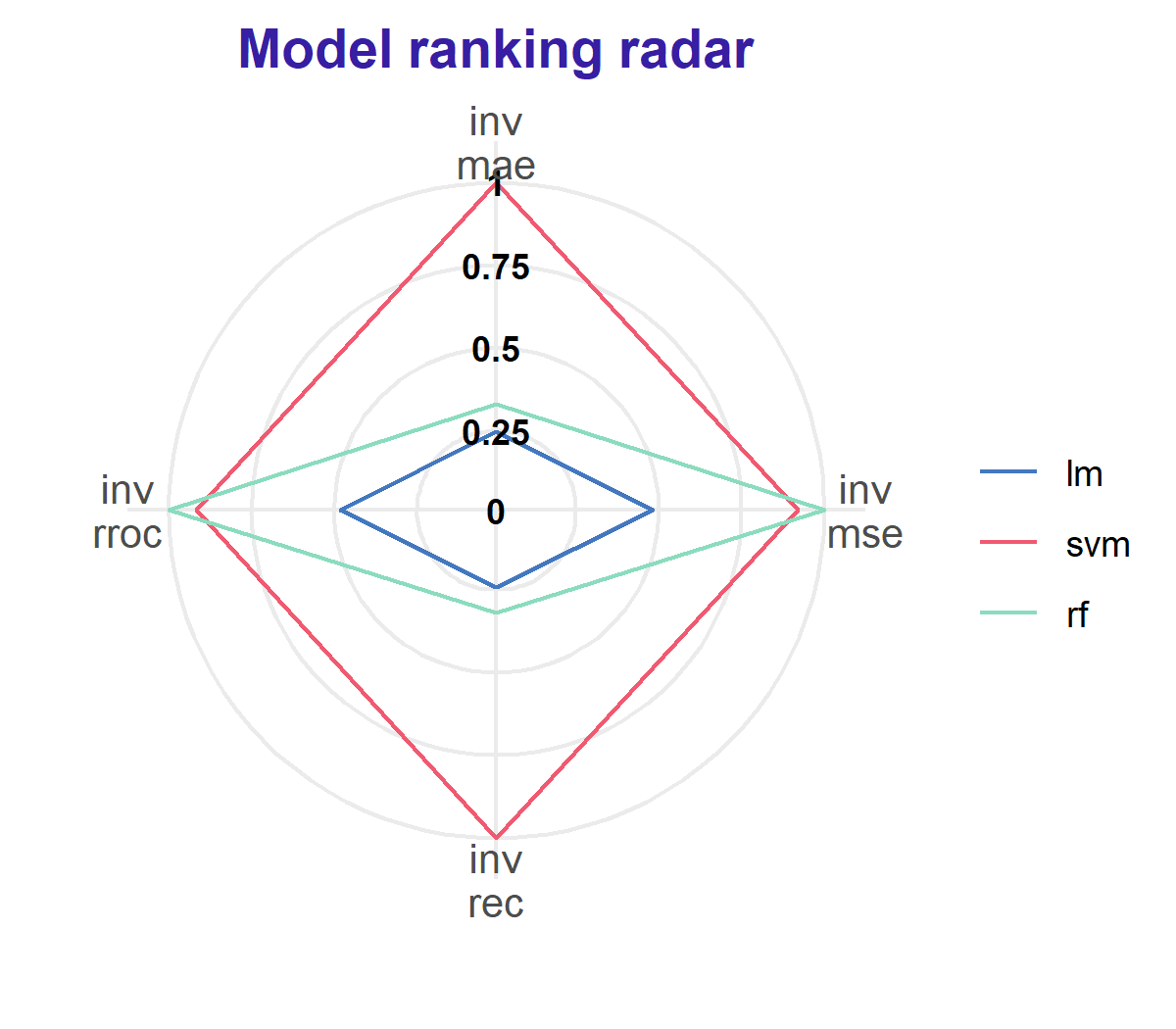}
      \caption[Use case: Model Ranking Plot and Predicted Response Plot]{ Model Ranking Plot for linear regression (blue), support vector regression (red), random forest (green). }
      \label{figure:1_uc}
\end{figure}

\begin{figure}[H]
\centering
      \includegraphics[width=0.75\textwidth]{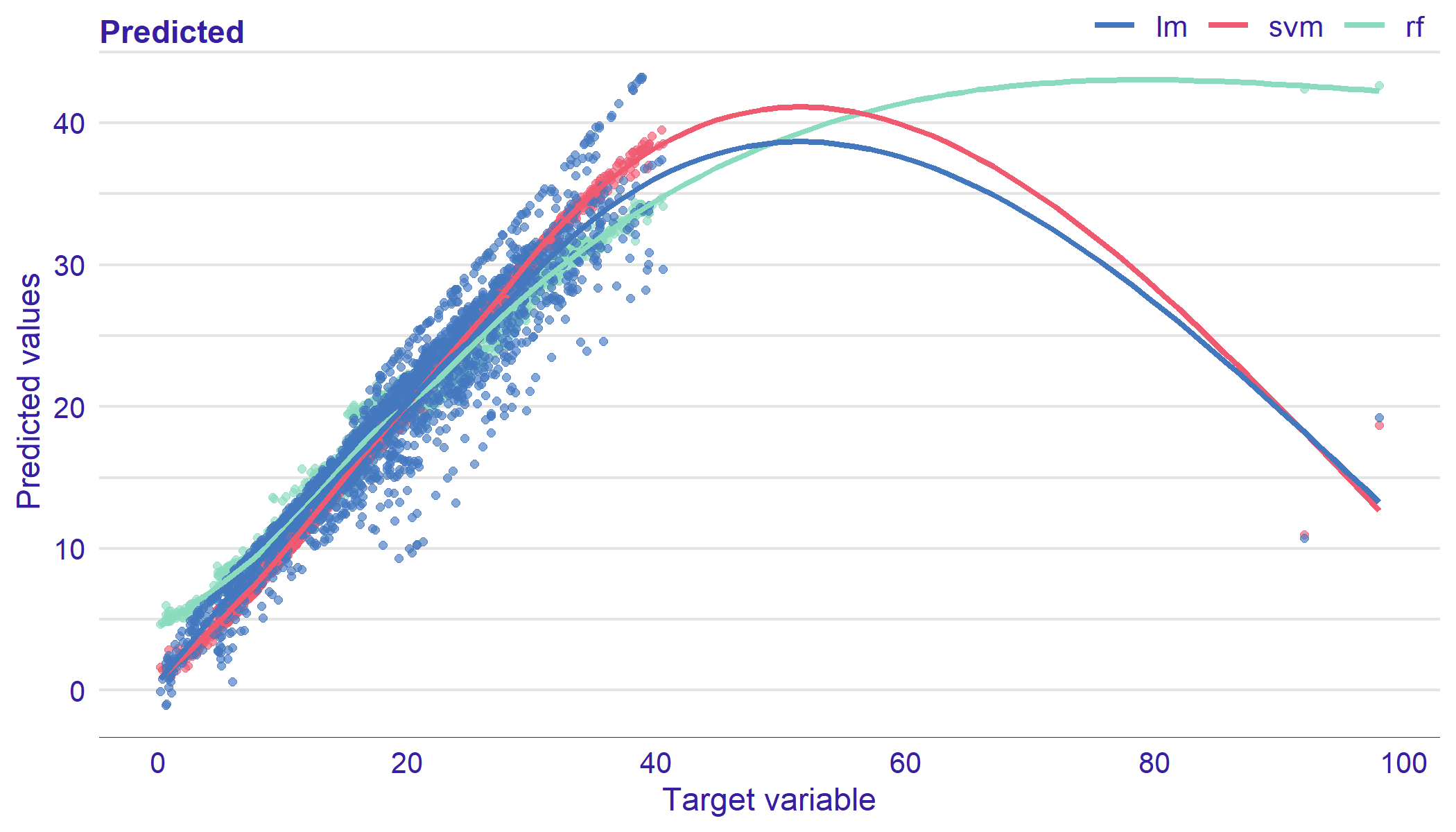}
      \caption[Use case: Model Ranking Plot and Predicted Response Plot]{Predicted Response~Plot for linear regression (blue), support vector regression (red), random forest (green).}
      \label{figure:1_uca}
\end{figure}

\begin{figure}[H]
 \centering
      \includegraphics[width=0.75\textwidth]{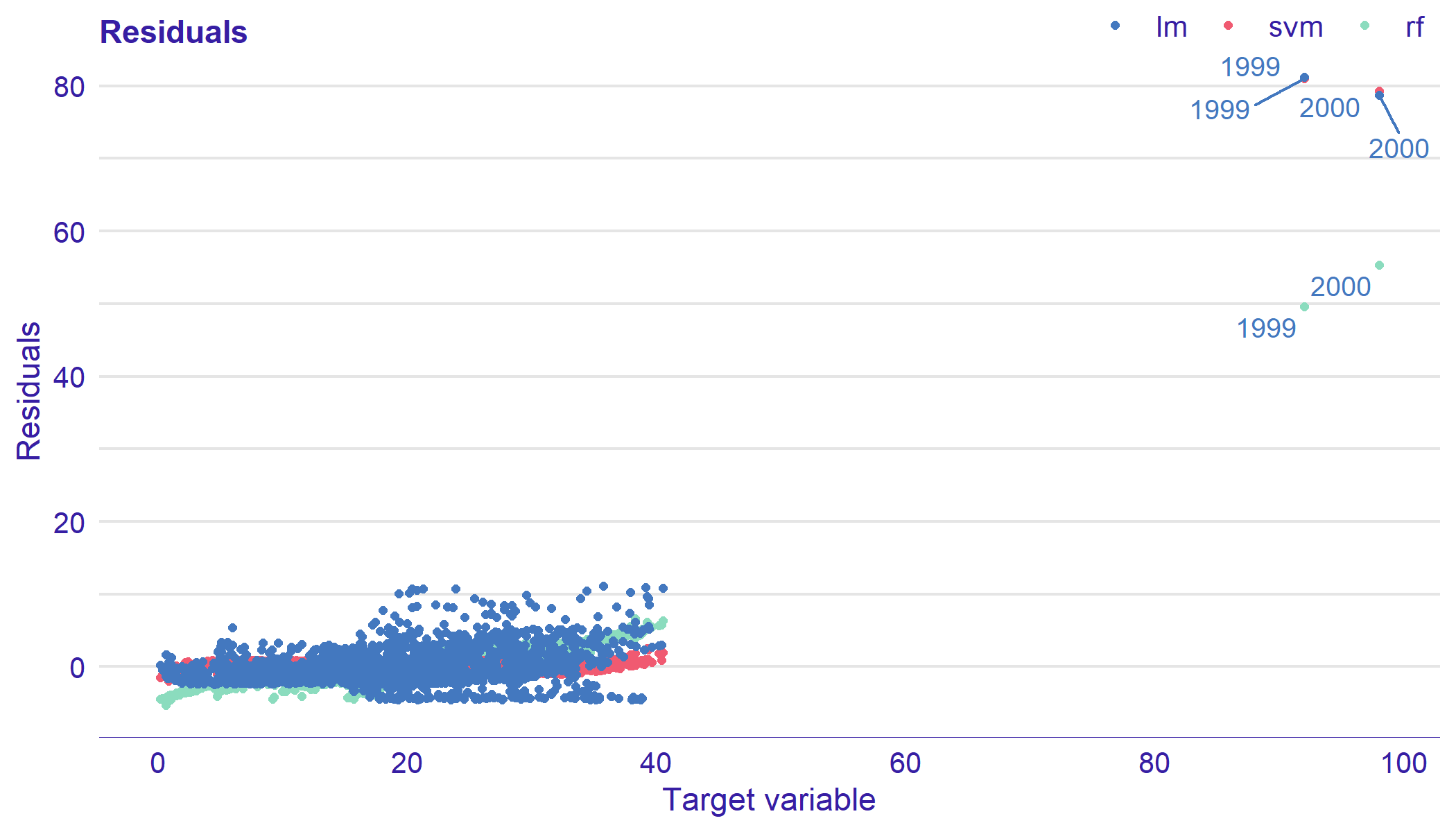}
      \caption[Use Case: Model PCA Plot and Residual plot]{Residual Plot for linear regression (blue), random forest (green), support vector regression (red). }
      \label{figure:2_uc}
\end{figure}

At first, we generate four diagnostic plots. They are: Model Ranking Plot, Predicted Response Plot, PCA of Models, and Residuals Plot. We present results in Figure~\ref{figure:1_uc}, Figure~\ref{figure:1_uca}, \mbox{and Figure~\ref{figure:2_uc}.}

\begin{example}
  lm_mp <- auditor::model_performance(au_lm)
  svm_mp <- auditor::model_performance(au_svm)
  rf_mp <- auditor::model_performance(au_rf)

  plot_radar(lm_mp, svm_mp, rf_mp)
\end{example}

\begin{example}
  lm_mr <- model_residual(au_lm)
  svm_mr <- model_residual(au_svm)
  rf_mr <- model_residual(au_rf)

  plot_prediction(lm_mr, svm_mr, rf_mr, smooth = TRUE)
\end{example}

\newpage

\begin{example}
  plot_residual(lm_mr, svm_mr, rf_mr, 
                variable = "_y_", nlabel = 6)
\end{example}

On the Model Ranking plot (see Figure~\ref{figure:1_uc}) we can see that random forest has the best performance in terms of $MSE$ and $RROC$ sores. Support vector regression is the best in terms of $MAE$ and $REC$ scores. Linear regression appears to have the worst performance in every aspect.
As none of the models performs best in terms of all measures, it is clear that Model Ranking plot is not enough to evaluate models. There is a need for further analysis of residuals.

Prediction Response plot (see Figure~\ref{figure:1_uca}) indicates that there are observations that may be outliers and have significant influence on model's structures. Most of the residuals of all models are arranged along a black line that shows the ideal trend. While, there is a group of points that clearly stands out from the rest. Further plots allow to take a closer look at these observations.

Residuals plot (see Figure~\ref{figure:2_uc}) confirms that there are 2 observations that have high influence on the structures of all models. Residuals plot with labeled points shows the numbers of outliers. These are $1999$-th and $2000$-th observations. The same that we added artificially.

\subsection{Removing identified outliers and models improvement}

In previous section, we identified two outliers in \code{auditorData}.
In this section, we remove those outliers, fit models to a new data set, and create \code{"model\_audit"} objects.
\begin{example}
  auditorData_clean <- auditorData[-c(1999, 2000), ]

  model_lm <- lm(y ~ ., auditorData_clean)
  set.seed(1994)
  model_rf <- randomForest(y~., auditorData_clean)
  model_svm <- svm(y ~ ., auditorData_clean)

  au_lm <- audit(model_lm, data = auditorData_clean, y = auditorData_clean$y)
  au_rf <- audit(model_rf, data = auditorData_clean, y = auditorData_clean$y, 
    label = "rf")
  au_svm <- audit(model_svm, data = auditorData_clean, y = auditorData_clean$y, 
    label = "svm")
\end{example}

\begin{figure}[H]
  \centering
      \includegraphics[width=0.5\textwidth]{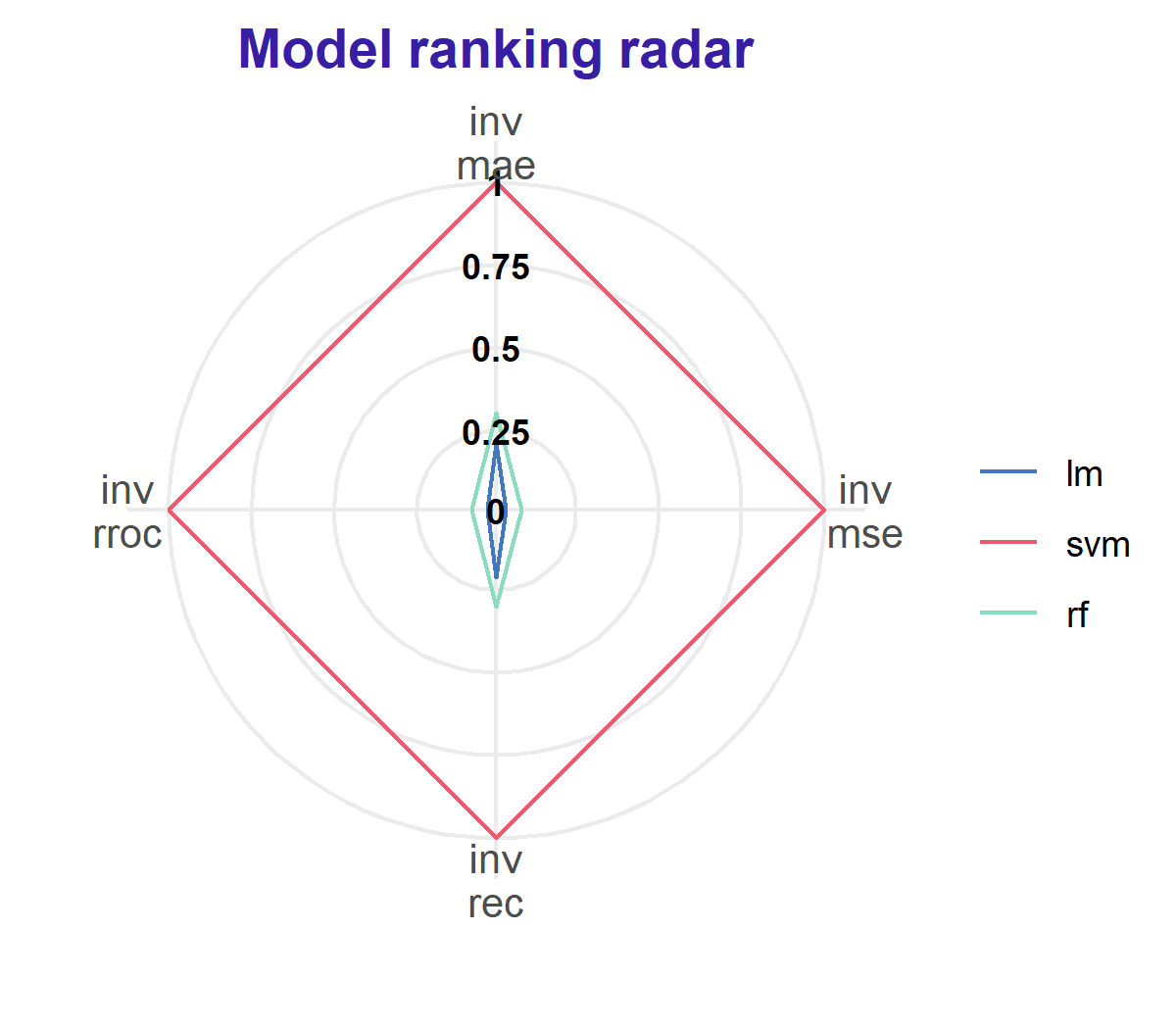}
      \caption[Use Case: New Model Ranking Plot and New Predicted Response Plot]{New Model Ranking Plot for linear regression (blue), random forest (green), support vector regression (red) fitted on new data set. }
      \label{figure:3_uc}
\end{figure}

\begin{figure}[H]
  \centering
      \includegraphics[width=0.75\textwidth]{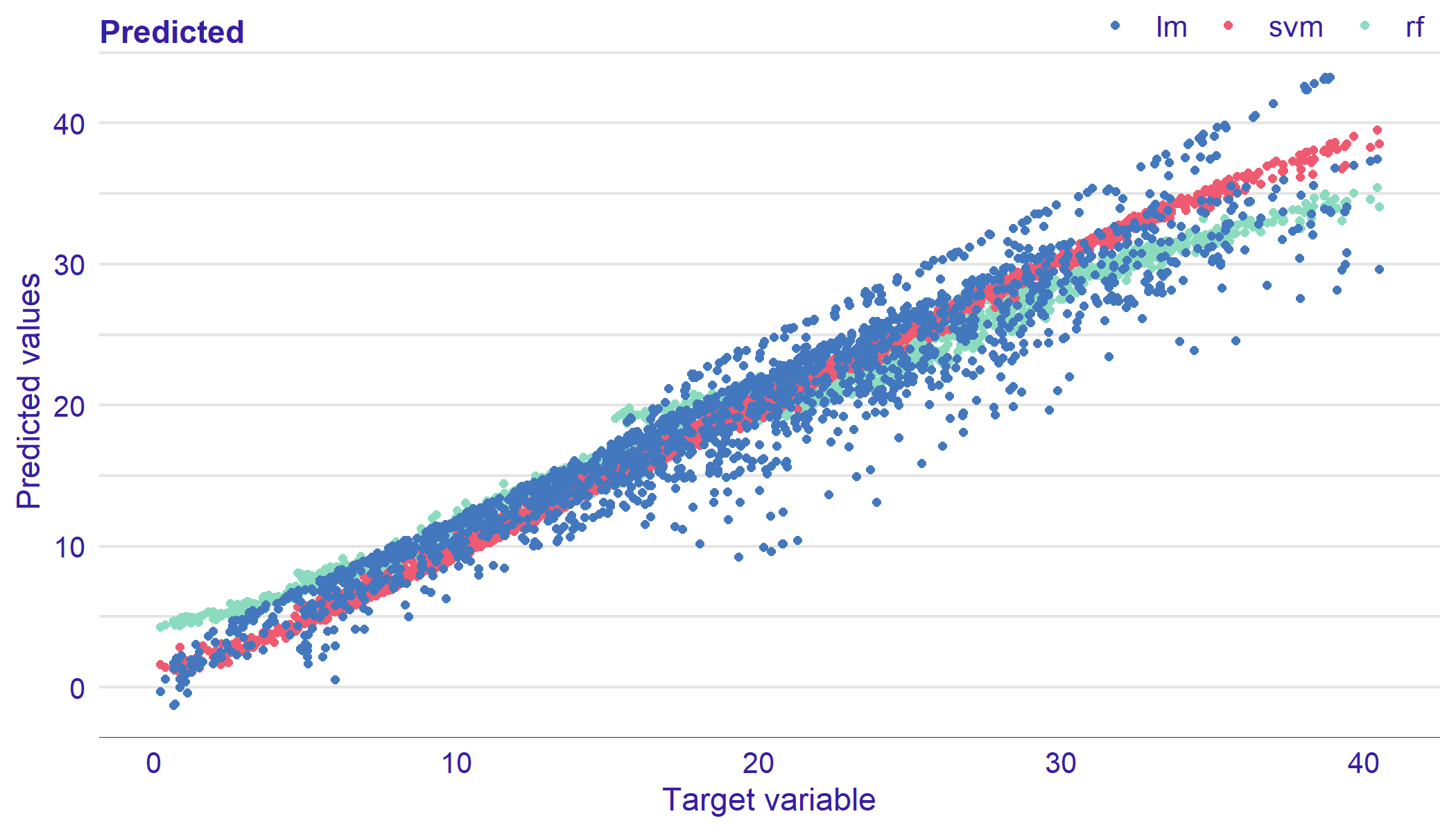}
      \caption[Use Case: New Model Ranking Plot and New Predicted Response Plot]{New Predicted Response Plot for linear regression (blue), random forest (green), support vector regression (red) fitted on new data set.}
      \label{figure:3_uca}
\end{figure}

We again generate Model Ranking (See~Figure \ref{figure:3_uc}) and Predicted Response Plots (See~Figure \ref{figure:3_uca}). Both of them \mbox{changed~noticeably.}

\begin{example}
  lm_mp_new <- model_performance(au_lm_new)
  svm_mp_new <- model_performance(au_svm_new)
  rf_mp_new <- model_performance(au_rf_new)

  plot_radar(lm_mp_new, svm_mp_new, rf_mp_new)
\end{example}

\begin{example}
  lm_mr_new <- model_residual(au_lm_new)
  svm_mr_new <- model_residual(au_svm_new)
  rf_mr_new <- model_residual(au_rf_new)

  plot_prediction(lm_mr_new, svm_mr_new, rf_mr_new)
\end{example}

This time, the Model Ranking plot clearly indicates that support vector regression performs best.
However, to better exploration of models, it is still worth to carry out a further audit. The~Predicted Response Plot shows that residuals of support vector regression are closest to the~diagonal line than residuals of other models. Residuals of linear model have the largest dispersion, while random forest residuals indicate over-prediction for small values of observed response and under-prediction for large values.

\subsection{Extended model audit}

In this section we take a closer look on the residuals. We generate Residual Boxplot and Two-sided ECDF plot for better comparison of models in aspect of residuals.
\begin{example}
  plot_residual_boxplot(lm_mr_new, svm_mr_new, rf_mr_new)
\end{example}
\begin{figure}[H]
      \includegraphics[width=0.9\textwidth]{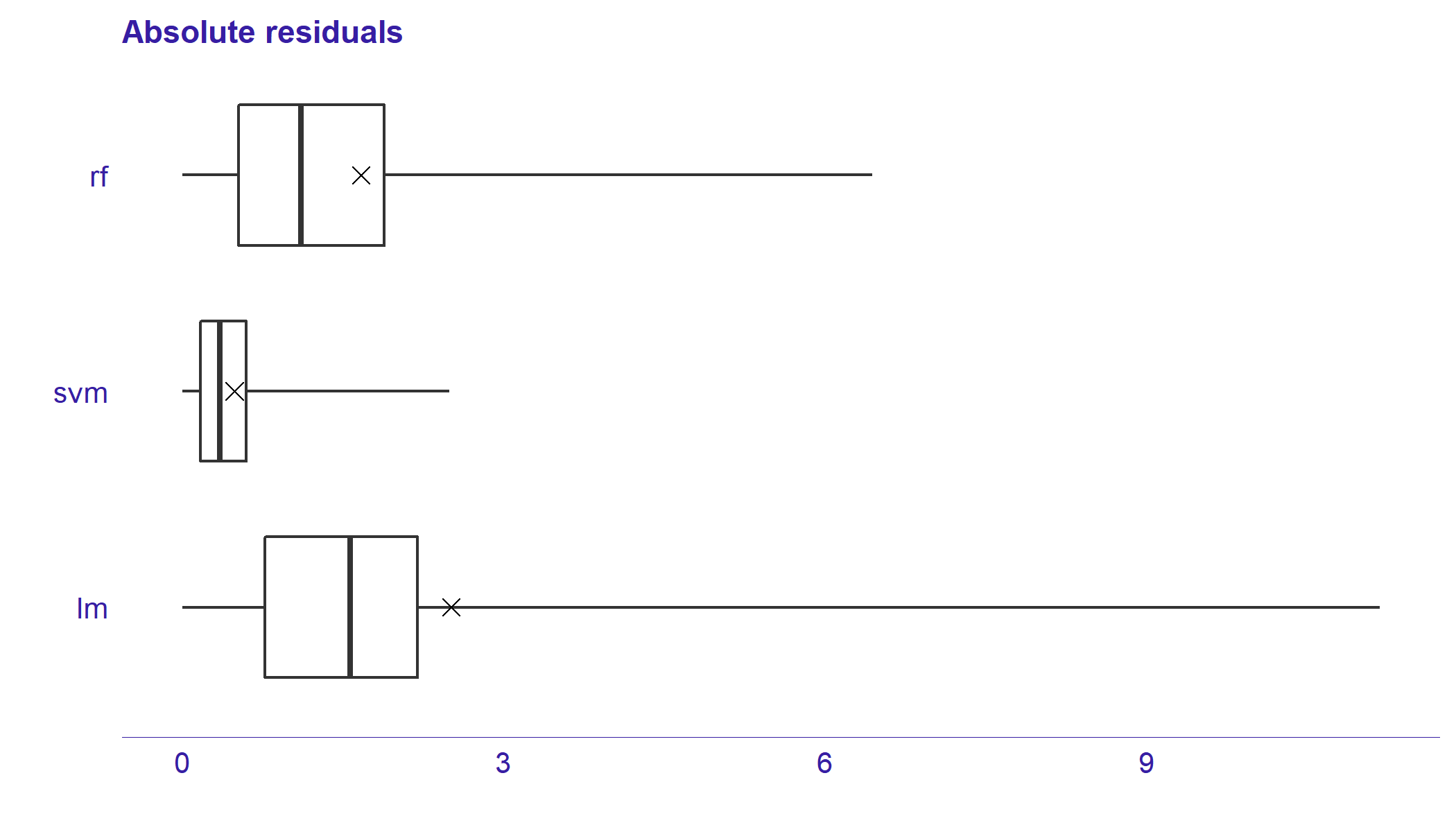}
      \caption[Use Case: Residual Boxplot]{Residual Boxplot for linear regression,  random forest, support vector regression.}
      \label{figure:4_uc}
\end{figure}

\begin{example}
  plot_tsecdf(lm_mr_new, svm_mr_new, rf_mr_new)
\end{example}
\begin{figure}[H]
      \includegraphics[width=0.9\textwidth]{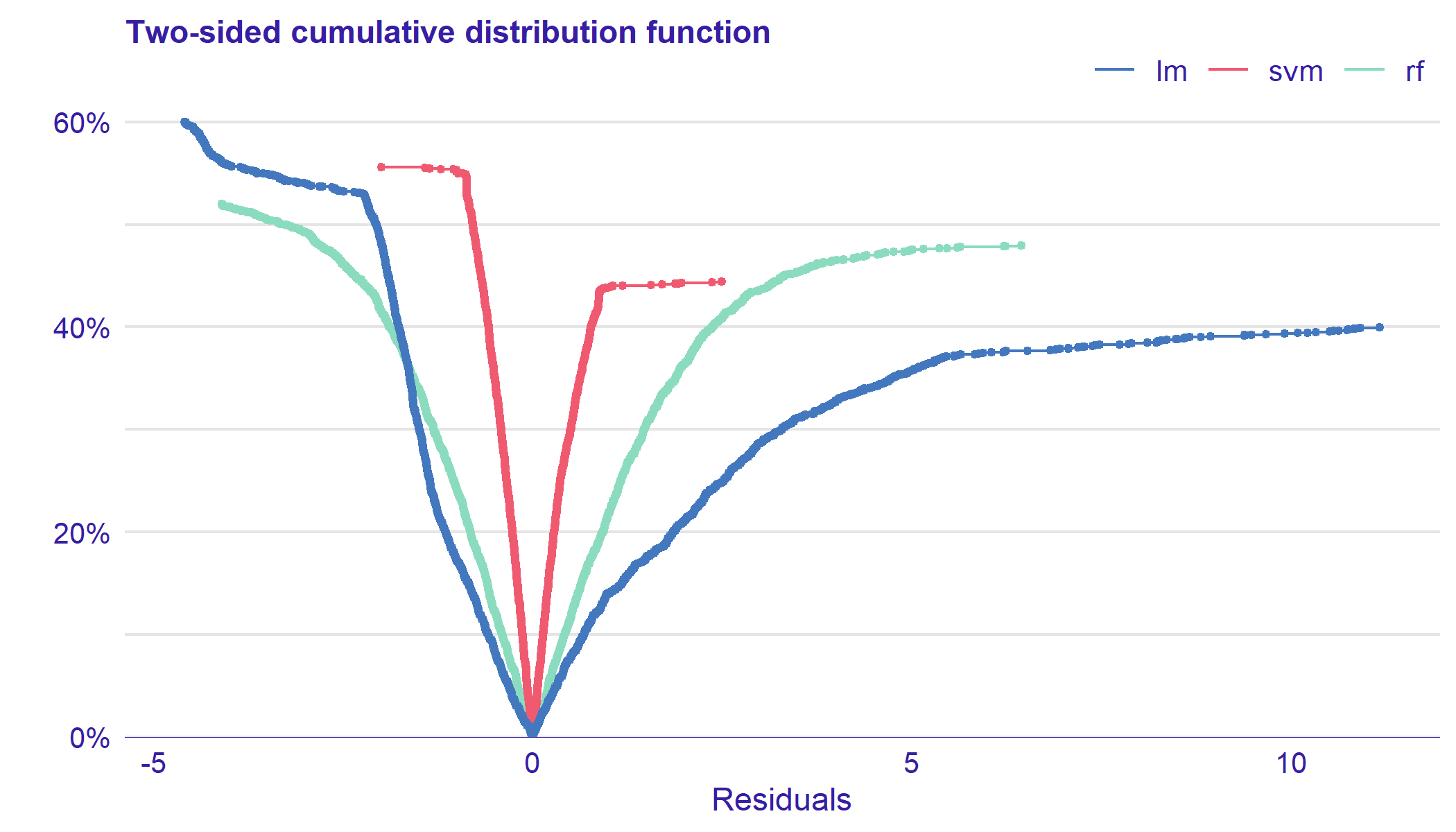}
      \caption[Use Case: Residual Boxplot]{Two-Sided ECDF Plot for linear regression (blue), random forest (green), support vector regression~(red). }
      \label{figure:4_uca}
\end{figure}

In Figure~\ref{figure:3_uc}, we noticed that support vector regression has the smallest residuals.
By analyzing boxplots (see Figure~\ref{figure:4_uc}), we can additionally see that random forest has smaller residuals than linear model. Two-sided ECDF plot (see Figure~\ref{figure:4_uca}) show that conclusions from boxplots are, in general, correct. In addition, we can see that growth of small negatives residuals for random forest and linear model is similar.

Now, we audit residuals due to the model variables. We use Residual Density to understand residual behaviour for different values of the $X4$ variable.

\newpage
\begin{example}
  plot_residual_density(lm_mr_new, svm_mr_new, rf_mr_new,
                        variable = "X4")
\end{example}
\begin{figure}[H]
\centering
      \includegraphics[width=0.7\textwidth]{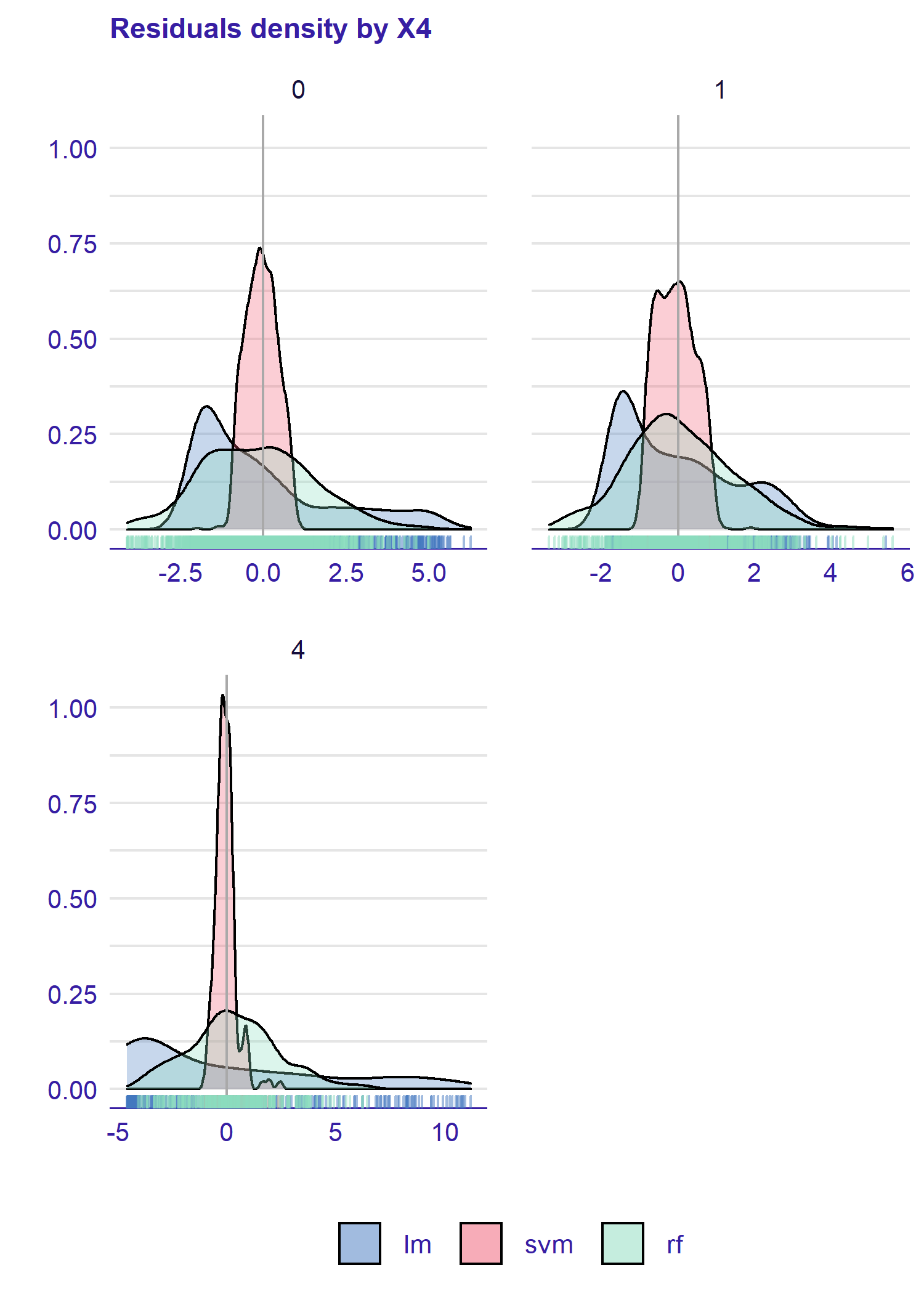}
      \caption[Use Case: Residual Density Plot]{Each panel of the Residual Density Plot  corresponds to one of the models. Colors indicates values of $X4$ variable.}
      \label{figure:5_uca}
\end{figure}
In the Figure~\ref{figure:5_uca}, we see that structure of residuals for random forest and support vector regression do not vary due to the value of $X4$ variable. In contrast, there are differences in the~shape of the densities for linear model's residuals. In the data set, we included the $X4$ variable in the interaction with $X1$. Due to its structure, the linear model do not catch this interaction.
We use Residuals Plot to examine behaviour of residuals due to second variable included in \mbox{the~interaction.}
\newpage
\begin{example}
  plot_residual(lm_mr_new, svm_mr_new, rf_mr_new,
                variable = "X1")

\end{example}
\begin{figure}[H]
\centering
      \includegraphics[width=0.75\textwidth]{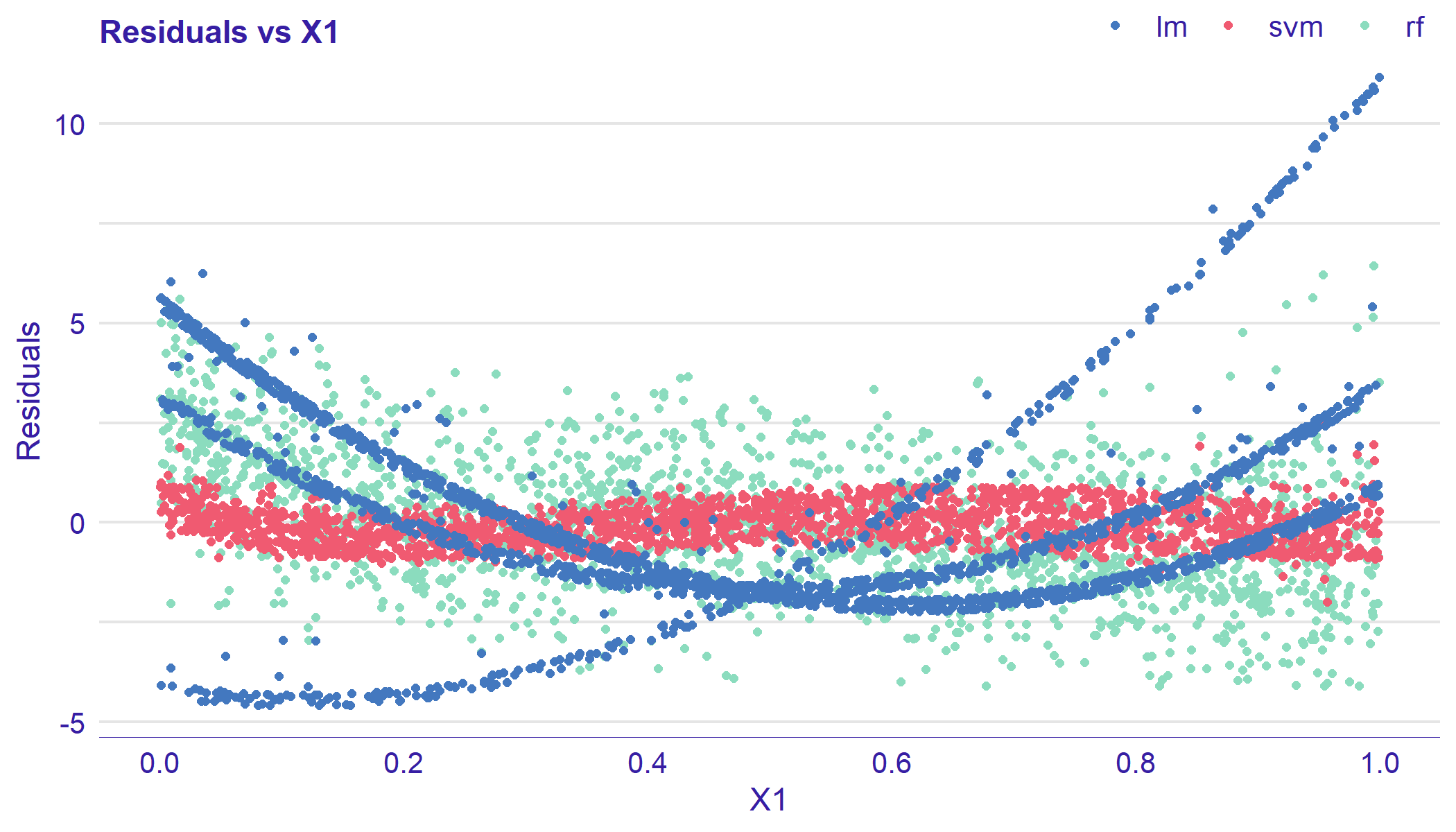}
      \caption[Use Case: Residuals Plot]{Residuals vs variable $X1$ Plot generated for random forest, support vector regression and linear model with ordinary least squares.}
      \label{figure:6_uc}
\end{figure}
In the Figure~\ref{figure:6_uc}, we see that residuals form three groups. Based on the previous analysis, we conclude that groups correspond to the levels of the $X4$ variable.

\subsection{Summary}
We fitted three different models to the data set \code{auditData}. We used Predicted Response, Model PCA and Residuals plots to identify two observations, that highly influenced model's structures. Then we refitted models on data without outliers. 
The audit showed that support vector regression has the best performance and we did not identify any serious problems with the structure of the residuals.

%% file: chapters/conclusion.tex
\section{Conclusion and future work} \label{conclusions}
In this article, we presented the \pkg{auditor} package and selected diagnostic scores and plots. We discussed the~existing methods of model validation and proposed new visual approaches. 
We also specified three objectives of model audit (see Section~\ref{introduction}), proposed relevant verification tools, and demonstrated their usage.
Model Ranking Plot and REC Curve enrich the information about model performance (Objective~1). Residual Boxplot, Residual Density, and Two-Sided ECDF Plots expand the knowledge about the distribution of residuals (Objective 3). What is more, the latter two tools allow for identification of outliers (Objective 2).

Aside from describing existing methods, we proposed new plots. They are Model Ranking Plot, Two-Sided ECDF Plot.

We implemented all the presented scores and plots in the \pkg{auditor} package for R. The included functions are based on a uniform grammar introduced in Figure~\ref{architecture}. 
Documentation and examples are available at~\url{https://modeloriented.github.io/auditor/}. The stable version of the package is on CRAN, the development version is on GitHub (\url{https://github.com/ModelOriented/auditor}).

In Section~\ref{use-case} we showed the use-case of the model audit with the \pkg{auditor}. We have presented a broad exploration and comparisons of three models by analyzing their residuals.

There are many potential areas for future work that we would like to explore, including more extensions of model-specific diagnostics to model-agnostic methods and residual-based methods for investigating interactions. 
Another potential aim would be to develop methods for local audit based on the diagnostics of a~model around a~single observation or a group of observations.